\documentclass{article}

\usepackage{arxiv}

\usepackage[utf8]{inputenc} 
\usepackage[T1]{fontenc}    
\usepackage{hyperref}       
\usepackage{url}            
\usepackage{booktabs}       
\usepackage{amsfonts}       
\usepackage{nicefrac}       
\usepackage{microtype}      
\usepackage{lipsum}
\usepackage{amsmath,amssymb}
\usepackage{braket}
\usepackage{graphicx}
\title{Phase resolved joint spectra tomography of a ring resonator photon pair source using a silicon photonic chip}

\author{Massimo Borghi
\\Quantum Engineering Technology Labs\\ H. H. Wills Physics Laboratory and Department of Electrical and Electronic Engineering\\University of Bristol, Bristol BS8 1FD, UK.\thanks{Present address: SM Optics s.r.l., Research Programs, Via John Fitzgerald Kennedy 2, 20871 Vimercate, Italy}
}

\begin{document}
\maketitle

\begin{abstract}
The exponential growth of photonic quantum technologies is driving the demand of tools for measuring the quality of their  information carriers. One of the most prominent is Stimulated Emission Tomography (SET), which uses classical coherent fields to measure the Joint Spectral Amplitude (JSA) of photon pairs with high speed and resolution. While the modulus of the JSA can be directly addressed from a single intensity measurement, the retrieval of the Joint Spectral Phase (JSP) is far more challenging and received minor attentions. However, a wide class of spontaneous sources of technological relevance, as chip integrated micro-resonators, have a JSP with a rich structure, that carries correlations hidden in the intensity domain. Here, using a compact and reconfigurable silicon photonic chip, it is measured for the first time the complex JSA of a micro-ring resonator photon pair source. The photonic circuit coherently excites the ring and a reference waveguide, and the interferogram formed by their stimulated fields is used to map the ring JSP through a novel phase reconstruction technique. This tool complements the traditionally bulky and sophisticated methods implemented so far, simultaneously minimizing the set of required resources.
\end{abstract}


\section{Introduction}
\label{sec:introduction}
One of the most appealing platforms for photonic quantum technolgies is integrated optics \cite{wang2019integrated,rudolph2017optimistic}. Since the first demonstrations more than a decade ago \cite{politi2008silica}, the complexity and scale of quantum optical circuits exponentially increased over the years \cite{harris2016large,qiang2018large,wang2018multidimensional,adcock2019programmable}. 
Irrespective of their application, key elements shared by every photonic circuit are photon sources. Silicon photonics architectures rely on Spontaneous Four Wave Mixing (SFWM) to probabilistically generate pairs of photons \cite{caspani2017integrated}. Waveguides sources constitutes the simplest example, emitting photons with high heralding efficiency but strong spectral correlations \cite{silverstone2014chip}. Complementarily, micro-resonators produce photons in almost pure states \cite{vernon2017truly,grassani2015micrometer,reimer2016generation}, but necessitates of tuning elements to ensure indistinguishability among independent sources \cite{faruque2018chip,silverstone2015qubit}. In parallel to source optimization, the developement of tools for their spectral characterization has known a burst since the introduction of Stimulated Emission Tomography \cite{liscidini2013stimulated}. SET allows to make predictions on SFWM (Spontaneous Parametric Down Conversion) based on the output of its classical counterpart, i.e., Stimulated Four Wave Mixing  (Difference Frequency Generation (DFG)) \cite{helt2012does}. Being the latter orders of magnitude more intense than the corresponding quantum process, integration times have been enormously decreased, and the resolution improved compared to spectrally resolved coincidence measurements \cite{zielnicki2018joint}. The Joint Spectral Amplitude is the function describing the spectral correlations of the photon pair \cite{caspani2017integrated}. 
So far, most of the applications of SET focused on the determination of the modulus of the JSA, referred as the Joint Spectral Intensity (JSI), since it is directly measurable from the power of the stimulated field. Examples includes silicon nanowires \cite{jizan2015bi}, AlGaAs ridge waveguides \cite{eckstein2014high}, optical fibers \cite{fang2014fast,erskine2018real}, All-Pass resonators \cite{grassani2016energy} and coupled rings \cite{kumar2014controlling}. A variation of SET, which implements Sum Frequency Generation (SFG), has been used to map the JSA of an array of evanescently coupled Lithium Niobate waveguides \cite{lenzini2018direct}. A method for the extraction of the complex JSA in multi-port devices, which implements DFG, has been proposed based on the eigenmode expansion of single channel excitations \cite{titchener2015generation}. SET has been also extended to others degrees of freedom, like space \cite{lenzini2018direct} and polarization \cite{fang2016multidimensional,rozema2015characterizing}. The Joint Spectral Phase (JSP) is of the same relevance of the JSI, but historically received minor attentions. The JSI itself is, however, an incomplete picture of the quantum state, since it hides the spectral correlations encoded in the phase domain \cite{jizan2016phase}. As an example, only the lower bound of the Schmidt number can be estimated if the JSP is not known \cite{eckstein2014high}. Beside quantum homodyne tomography of two photon states \cite{ren2012analysis,beduini2014interferometric}, which suffers the same time and resolution issues of spectrally resolved coincidence measurements, phase resolved applications of SET have been so far limited to in-line sources as waveguides \cite{jizan2016phase,avenhaus2014time} or cold atomic ensembles \cite{park2017measuring}. In all these works, the Pump, the Seed and the reference beams are all carved from the same laser source to guarantee mutual coherence, and fiber based interferometers are used to extract the JSP. 

\noindent Here, a novel technique which allows to measure the complex JSA of a double bus, integrated silicon ring resonator, is proposed and experimentally validated. The method exploits the on-chip interference between the stimulated field of the ring, carrying the information on the JSA, and the one of a coherently pumped reference waveguide. Outside to the hypothesis of SET, the stimulating Seed is not mimicking the asymptotic output field of the corresponding photon of the pair \cite{liscidini2012asymptotic}. This makes the excitation scheme trivial, since both the Seed and the Pump lasers are directly coupled into the same port of the resonator, but comes at the expense of determining the transfer function of the device. However, the the integrated circuit is designed to to perform this task without additional complexity. Phase resolved tomography is completely implemented on a chip (an external filter is solely used to increase the spectral resolution), without the use of any auxiliary reference laser, since this is naturally provided by the broadband emission of the reference waveguide. Thermal phase shifters allow the circuit to perform different tasks, as addressing the individual JSI of the ring and of the waveguide, manipulating their interference, and probing the device transfer function.  Through an hamiltonian treatement  \cite{liscidini2012asymptotic}, it is rigorously proved that the scheme can be applied to a single resonator with an arbitrary number of channels, even if they are not physically accessible, as the ones associated to losses \cite{vernon2015spontaneous}. 


\section{Stimulated emission tomography on an Add-Drop resonator}
\label{sec:theory}
\begin{figure*}[t!]
\centering
\includegraphics[scale = 0.46]{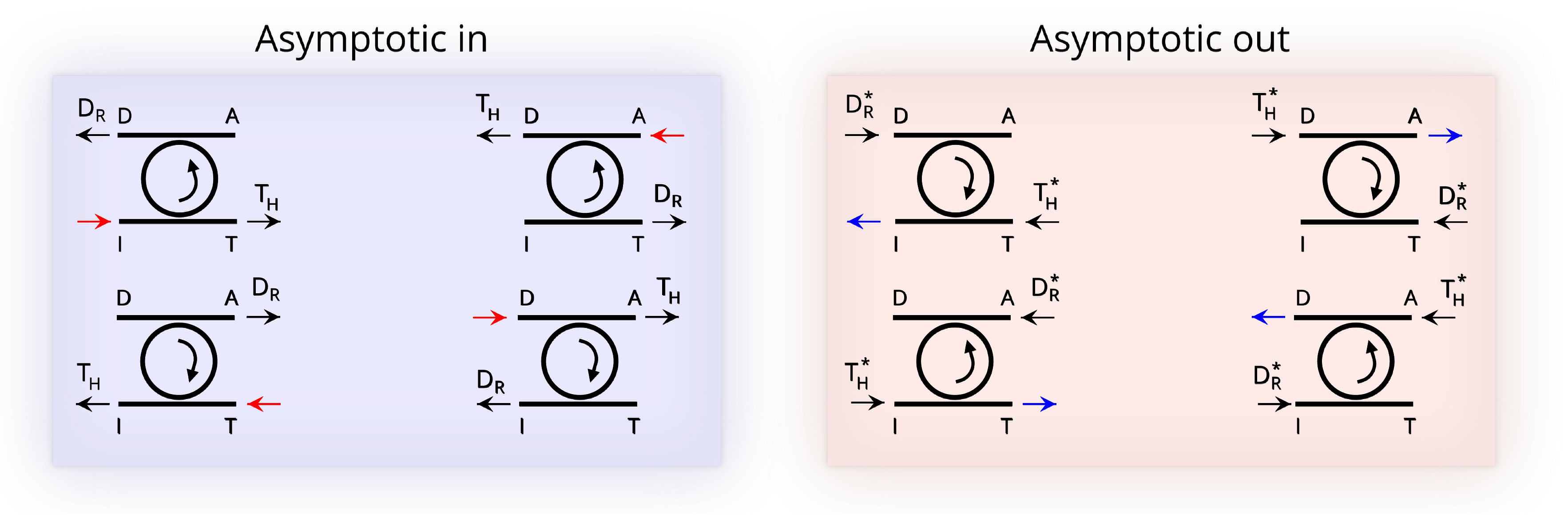}
\caption{The four asymptotic input (blue box) and output (red box) fields of the system. Arrows indicate the direction of energy flow. The asymptotic fields are labelled according to the ports pointed by their corresponding colored arrow. The (wavelength dependent) complex amplitude associated to each arrow is labeled as $T_H$ if it refers to the Through transfer function, or $D_R$ if it refers to the Drop one.}
\label{fig:1}
\end{figure*}

One of the key hypothesis of SET is that the Seed laser, with wavevector $k_s$, must be coupled in the same asymptotic output field of the corresponding photon of the pair \cite{liscidini2013stimulated}. In this case, the amplitude $\gamma(k_i)$ of the stimulated field at wavevector $k_i$ is directly proportional to the JSA $\phi(k_s,k_i)$. The aim of this section is to derive a more general relation between $\phi$ and $\gamma(k_i)$, which holds even when the Seed is not an asymptotic output field (as in this work), and from this to define an alternative method for recovering $\phi$.   
The specific case of interest is of an Add-Drop (double bus) resonator with four channels, which will be labelled as Input, Through, Add and Drop (respectively I,T,A and D in Fig.\ref{fig:1}). As a first step, the device will be considered as lossless, a condition that will be relaxed later. The Pump and the Seed lasers are coupled to the Input port, and the intensity of the stimulated field (the Idler) is monitored in either  the Drop or the Through bus waveguide. The nonlinear hamiltonian $H_{\textup{nl}}$ responsible for SFWM is \cite{park2017measuring}:
\begin{equation}
H_{\textup{nl}}  = -\sum_{xy}\int S_{xy}(k_{p_1},k_{p_2},k_{i},k_{s})a_{I,k_{p_1}}a_{I,k_{p_2}}b_{x,k_{i}}^{\dagger}b_{y,k_{s}}^{\dagger}dk_{p_1}dk_{p_2}dk_{s}dk_{i} + \textrm{h.c.}
\label{eq:1}
\end{equation}
where $x(y) = \{T,D\}$ labels the channels and $k$ the wavevectors of the Pump, the Seed and the Idler. Operators $a_{I}$ and $b_{j}^{\dagger}$ respectively annihilate and create the asymptotic input field of channel $I$, and the asymptotic-output field of channel $j$. A sketch of these states is shown in Fig.\ref{fig:1}. The function $S_{xy}$ contains the spatial overlap between the four asymptotic fields as well as the phase matching function \cite{yang2008spontaneous}. At any time $t$, the state of the system $\ket{\psi(t)}$ can be written as a tensor product of the Pump, the Signal and the Idler coherent beams:
\begin{equation}
\ket{\psi(t)} = D_{p,I}(t)D_{s,T}(t)D_{s,D}(t)D_{i,T}(t)D_{i,D}(t)\ket{\textup{mod}(t)}
\label{eq:2}
\end{equation}
where $D_{mn}$ is the displacement operator associated to the beam $m=\{p,s,i\}$ in the channel $n=\{I,T,D\}$. In the case of the Signal and the Idler beams, these are defined as:
\begin{equation}
D_{m,n}(t) = \exp \left( \int{\gamma_{mn}(k,t)b^{\dagger}_{n,k}dk} - \textrm{h.c.} \right) \label{eq:3}
\end{equation}
where $\gamma_{mn}(k,t)$ represents the instantaneous wavevector distribution  of beam $m$ in channel $n$ at time $t$. These are assumed to be peaked around the central wavevectors $(k_{s0},k_{i0})$, which correspond to two distinct and not overlapping resonance orders. $D_{p,I}$ is similarly defined, but uses asymptotic input field creation operators. The Seed laser is assumed monochromatic, so as $\gamma_{sT(D)}(k_s)=\gamma_{sT(D)}(k_{s0})\delta(k_s-k_{s0})$. At time $t_0\rightarrow -\infty$, an initial state $\ket{\psi(t_0)}$ is constructed such as the Pump and the Seed  have not entered yet in the nonlinear region, so there is no stimulated field ($\gamma_{in}(k,t_0)=0$) and $\ket{\textup{mod}(t_0)}=\ket{0}$.  When this state is evolved through $H_{\textrm{nl}}$ to a time $t_f\rightarrow \infty$, such that all the energy has left the nonlinear region, we have, in the undepleted Pump and Seed approximation (derivation in Appendix A), that \mbox{$\
\ket{\psi(\infty)} = D_{p,I}(\infty)D_{s,T}(\infty)D_{s,D}(\infty)D_{i,T}(\infty)D_{i,D}(\infty)\ket{\textup{mod}(\infty)}$}, where:
\begin{equation}
\ket{\textup{mod}(\infty)}= \ket{0}+\sqrt{p}\sum_{xy} \frac{\phi_{xy}(k_s,k_i)}{\sqrt{p_{xy}}}b_{x,k_i}^{\dagger}b_{y,k_s}^{\dagger}\ket{0} dk_s dk_i
\label{eq:4}
\end{equation}
in which $D_{p,I}(\infty)\sim D_{p,I}(t_0)$ and $D_{s,j}(\infty) \sim D_{s,j}(t_0)$ ($j=\{T,D\}$). The quantity $p_{xy}$ is the probability of finding a photon of the pair in channel $x$ and its partner in channel $y$, while $p=\sum_{xy}p_{xy}$ is the overall probability of generating the pair. The functions $\phi_{xy}$ are defined as:
\begin{equation}
\phi_{xy}=  \frac{2\pi i}{\hbar\sqrt{p_{xy}}}\int{S_{xy}(k_{p_1},k_{p_1}+k_s-k_i,k_i,k_s)} \gamma_{pI}(k_{p_1})\gamma_{pI}(k_{p_1}+k_s-k_i)dk_{p_1}
\label{eq:5}
\end{equation}
The quantity $|\phi_{xy}(k_s,k_i)|^2$ is the normalized density probability of generating a photon with wavevector $k_i$ in the asymptotic output field $x$, and a photon with wavevector $k_s$ in the asymptotic output field $y$, so they represent JSAs. A complete tomography of the Add-Drop would require to determine the four JSAs. However, as it will be shown later, these are trivially related, and the knowledge of any of the JSAs suffices to determine them all.
As indicated in Eq.\ref{eq:2}, the state of the Idler is the product of two orthogonal coherent states, one associated to the asymptotic output field of the Drop and the other of the Through. The wavevector distribution of these two states are (see Appendix A) :
\begin{equation}
\begin{aligned}
\gamma_{iT}(k_i) = & \sqrt{p_{TT}} \phi_{TT}(k_{s0},k_i)\gamma_{sT}(k_{s0})^*+\sqrt{p_{TD}}\phi_{TD}(k_{s0},k_i)\gamma_{sD}(k_{s0})^*  \\
\gamma_{iD}(k_i) = & \sqrt{p_{DD}} \phi_{DD}(k_{s0},k_i)\gamma_{sD}(k_{s0})^*+\sqrt{p_{DT}}\phi_{DT}(k_{s0},k_i)\gamma_{sT}(k_{s0})^* 
\end{aligned}
\label{eq:6}
\end{equation}
The result of Eq.\ref{eq:6} tells that the JSAs are actually the kernels which relate the amplitude of the stimulated field to the ones of the asymptotic-output fields of the Seed. If the latter is exactly mimicking one of them, i.e., $\gamma_{sy}(k_{s0})=0$ and $\gamma_{sy'}(k_{s0})\neq0$ with $y \neq y'$, then the amplitude of the stimulated field in channel $x$ is directly proportional to the wavefunction $\phi_{xy'}$. However, in most of the cases it is impractical to engineer the Seed such that this condition is realized. Even in the simple example of a lossless All-Pass ring, the reconstruction of the output state associated to the Through port requires to inject the Seed laser at the Input with a wavelength dependent amplitude $T_H(\lambda_s)^*$, where $T_H$ is the complex transfer function of the Through port of the resonator \cite{liscidini2012asymptotic}. 
A way to overcome this complexity is to seed only one port, as the Input, and to exploit the relations in Eq.\ref{eq:6}. If the two coupling regions are equal, the phase matching functions $S_{xy}$ do not depend on the channel combination, so that $\phi_{xy}=\phi$.
 It is easier to express the state of the Seed in terms of asymptotic inputs, using the general input-output relations derived in \cite{liscidini2012asymptotic}, which give (see Appendix A):
\begin{equation}
\begin{aligned}
\gamma_{sT}(k_{s0}) & = T_H(\lambda_{s0}) \gamma_{sI}(k_{s0})+D_R(\lambda_{s0}) \gamma_{sA}(k_{s0}) \\
\gamma_{sD}(k_{s0})& = D_R(\lambda_{s0}) \gamma_{sI}(k_{s0})+T_H(\lambda_{s0}) \gamma_{sA}(k_{s0}) \\
\end{aligned} 
\label{eq:7}
\end{equation}
When Eq.\ref{eq:7} is inserted in Eq.\ref{eq:6}, \mbox{$\gamma_{iT(D)}(k_{i}) = \sqrt{\frac{p}{p_{TT}}} \gamma_{sI}(k_{s0})^* \phi(k_s,k_i) (T_H+D_R)^*$}, where it is used the fact that $\gamma_{sA}(k_{s0})=0$ since the Seed is injected only at the Input port. 
From Temporal Coupled Mode Theory (TCMT) applied to a weakly coupled resonator \cite{li2016backscattering}, it is possibile to prove (Appendix B) that $(T_H+D_R)^*=(\textup{FE})^*/\textup{FE}$, where $\textrm{FE}$ is the internal field enhancement of the resonator. This gives the final expression:

\begin{equation}
\gamma_{iT(D)}(k_i)  = \sqrt{p}\gamma_{sI}(k_{s0})^* |\phi(k_{s0},k_i)|e^{(\theta_{\phi}(k_{s0},k_i)+2\theta_{\textup{FE}}(k_{s0}))}
\label{eq:8}
\end{equation}
where \mbox{$\theta_{\phi}(k_{s0},k_i) =\textrm{Arg}(\phi(k_{s0},k_i))$} is the JSP and \mbox{$\theta_{\textup{FE}}(k_{s0}) = \textrm{Arg}(\textrm{FE}(k_{s0}))$}. 
As explicitly derived in Appendix A, this result holds even for a resonator with an arbitrary number of channels, having different coupling rates with the ring, and is consistent with the classical result predicted by TCMT. As expected, since the Seed is not injected in an asymptotic output field, the amplitude of the stimulated field and the complex JSA are no more proportional. Worth to note that $|\gamma_{sT(D)}(k_{s0})|\propto |\phi|$, so the usual SET procedure can still be applied to determine the JSI only by monitoring the intensity of the stimulated field.\\ 
\noindent It is important to stress the assumptions behind Eq.\ref{eq:8}, to show the limitations of this approach. As explicitly derived in Appendix A, provided that the transfer function of the device is accessible, and that all the asymptotic fields within the nonlinear region are trivially related through some constants, it is always possible to express the different JSAs $\phi_{xy}$ as $\phi_{xy}=\kappa_{xy}\phi$, where $\phi$ is a reference wavefunction and $\kappa_{xy}$ is a (frequency dependent) factor. This allows to factorize the terms $\phi_{xy}$ in Eq.\ref{eq:6} and use the relations in Eq.\ref{eq:7} to express the amplitude of the stimulated field as $\propto \phi(k_{s0},k_i)Q(k_{s0})$, where $Q(k_{s0})$ is a linear combination of the transfer functions of the channels. In our specific case, $\kappa_{xy}=1$ and $Q(k_{s0})=(T_H(k_{s0})+D_R(k_{s0}))^*$. Hence, since $Q(k_{s0})$ must be known, the proposed method can not be applied to "black boxes", but only to those systems whose transfer function has been queried in advance. An example of system where this method fails is a coupled resonator chain, where the asymptotic fields inside the different resonators could have complex nonlinear relations between each other. 

\subsection{JSA reconstruction}
In order to determine the JSA outside the hypothesis of SET, the complex amplitude $\gamma_{iT(D)}(k_{si})$ of the stimulated field and the phase $\theta_{\textup{FE}}$ of the field enhancement have to be measured. The strategy that will be adopted in the following is to coherently excite the resonator and a reference waveguide, and to make their stimulated fields to interfere in order to address their relative phase. The reference source should have an almost flat amplitude profile over the bandwidth of the resonator, such that their relative phase is, up to an overall constant factor, following the same variations of the one of the ring. Provided that the Pump does not carry significant chirp, a waveguide source meets this requirement, since its FWM bandwidth can be easily made to extend by more than $30 \,\textrm{nm}$ \cite{borghi2017nonlinear}.
Using a similar strategy, $\theta_{\textup{FE}}$ can be extracted. Since the field circulating in the resonator can not be directly accessed, this is circumvented by measuring  the complex transfer function of the Through port, and by using the TCMT relation $\textrm{FE}=-i\sqrt{\frac{\tau_e}{2\tau_{\textup{rt}}}}(T_H(\lambda)-1)$ (where $\tau_e$ is the extrinsic photon lifetime associated to loss into the bus waveguide and $\tau_{\textup{rt}}$ is the roundtrip time of light into the cavity) to extract the complex field enhancement. The modulus of $T_H$ is given by the intensity of the light transmitted by the resonator, while the phase is measured relative to the one of the reference source. 

\section{Device and experimental setup}
\label{sec:experiment}
\begin{figure*}[t!]
\centering
\includegraphics[scale = 0.36]{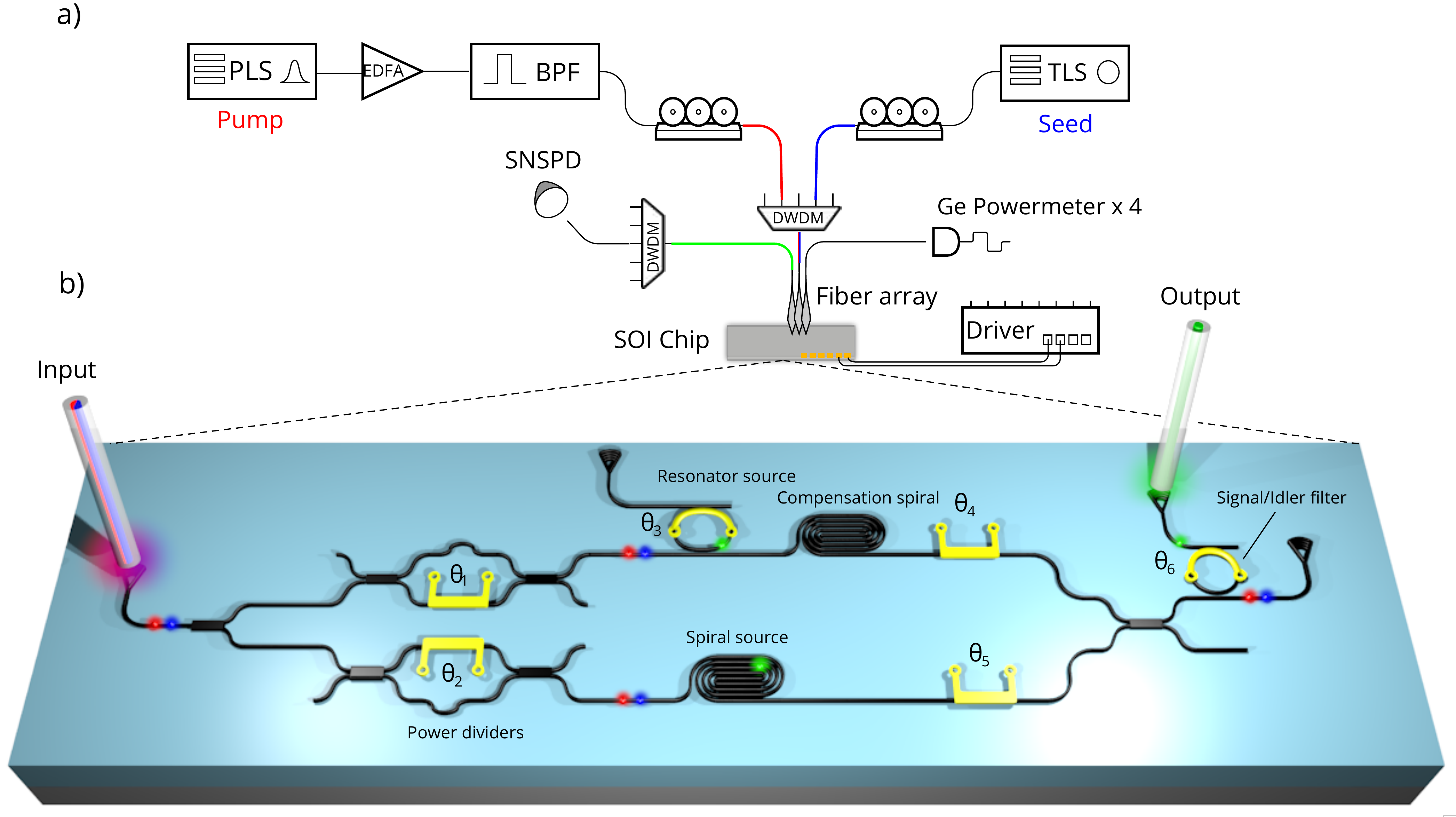}
\caption{(a) Sketch of the experimental setup. The Pump and the Seed lasers are respectively indicated with red and blue colors, while the stimulated field is indicated in green. PLS: Pulsed Laser Source, TLS: Tunable Laser Source, BPF: Band Pass Filter, FPC: Fiber Polarization Controller. (b) Layout of the chip. Waveguides are shown in black, while heaters are indicated in yellow.}
\label{fig:1_1}
\end{figure*}

The setup for the JSA reconstruction is sketched in Fig.\ref{fig:1_1}(a). The Pump is a femtosecond pulsed laser (Pritel), tuned at the resonance wavelength $\lambda_{m_p} = 1553.5\,\textrm{nm}$ and with a repetition rate of $50\,\textrm{MHz}$. The spectral width is set to $250\,\textrm{pm}$ by using a variable bandwidth tunable filter (Yenista XTA-50). The Pump is combined to a CW laser (Yenista TS100-HP) using a $200\,\textrm{GHz}$ commercial Dense Wavelength Division Multiplexing module (DWDM, Opneti), which also cleans the background noise of the laser. The polarization is set to TE by using Fiber Polarization Controllers. Light is injected and collected to and from the chip using a $16$ channel Fiber Array and grating couplers. After loss calibration, an average Pump power of $-8.5\,\textrm{dBm}$ and a Seed power of $-4.5\,\textrm{dBm}$ have been estimated at the Input waveguide of the circuit.
The device, sketched in Fig.\ref{fig:1_1}(b), is patterned on a $220$ nm SOI wafer using Electron-Beam lithography (EBL) from the Applied Nanotools foundry \cite{ant}. Single mode waveguides have a cross section of $500\times220\,\textrm{nm}^2$
and lie on a $2\,\mu\textrm{m}$ thick Buried Oxide (BOX) layer. A $2.2\,\mu\textrm{m}$ thick Silica layer is deposited on the top of the waveguides, which provides optical isolation from the heater layer. 
 All the thermal phase shifters can be simultaneously controlled by an external multi-channel current driver.\\
\noindent The circuit can be divided into three stages. In the first, depending on the choice of the phases $\theta_{1(2)}$, the Pump and the Seed lasers can be directed to the upper and/or lower arms of the device. The routing is achieved by asymmetric Mach-Zehnder interferometers (aMZI) with an FSR of $800\,\textrm{GHz}$ and an Extinction Ratio (ER) of $-35\,\textrm{dB}$. 
In this way, in the second stage, stimulated FWM can isolately occur in the ring resonator, or in the reference spiral, or simultaneously in both of them. The resonator source is a double bus ring of mean radius $13.87\,\mu\textrm{m}$, a measured linewidth of $250\,\textrm{nm}$ (quality factor $Q=6200$), FSR of $800\,\textrm{GHz}$ and ER of $-13\,\textrm{dB}$. The reference source is a spiral waveguide with a length of $L=2.35\,\textrm{mm}$, which has been engineered to have a FWM bandwidth of more than $30\,\textrm{nm}$ and a comparable brightness to the one of the resonator. 
A spiral has been added after the resonator to compensate the path length mismatch between the upper and lower arm of the circuit. In this way, the stimulated fields, which are manipulated in the third stage, experience the same optical path from the sources to the final beamsplitter (based on a MultiMode Interference device). Through the tuning of $\Delta\theta = \theta_4-\theta_5$, the relative phase of the two stimulated fields can be varied. An on-chip filter for the stimulated radiation is used, which is based on an Add-Drop ring resonator. This has an FSR of $1200\,\textrm{GHz}$, and due to unexpectedly high bending losses, the  ER is only $-2\,\textrm{dB}$ and the linewidth $140\,\textrm{pm}$. The phase $\theta_6$ sweeps the filter wavelength across the Idler  resonance order $m_i$. The stimulated field at the output of the chip is isolated from the Pump and the Seed laser by using a DWDM, and directed to a Superconducting Nanowire Single Photon Detector (SNSPD), operating at $85\%$ detection efficiency and with a dark count level $<200\,\textrm{Hz}$. 
Optionally, light can be directed to an off-chip tunable filter (Yenista XTA-50) to increase the spectral resolution to $50\,\textrm{pm}$ (Full With at Half Maximum (FWHM)). 
\section{Measure of the JSI using the on-chip filter}
\label{sec:measureJSI}
\begin{figure}[h!]
\centering
\includegraphics[scale = 0.7]{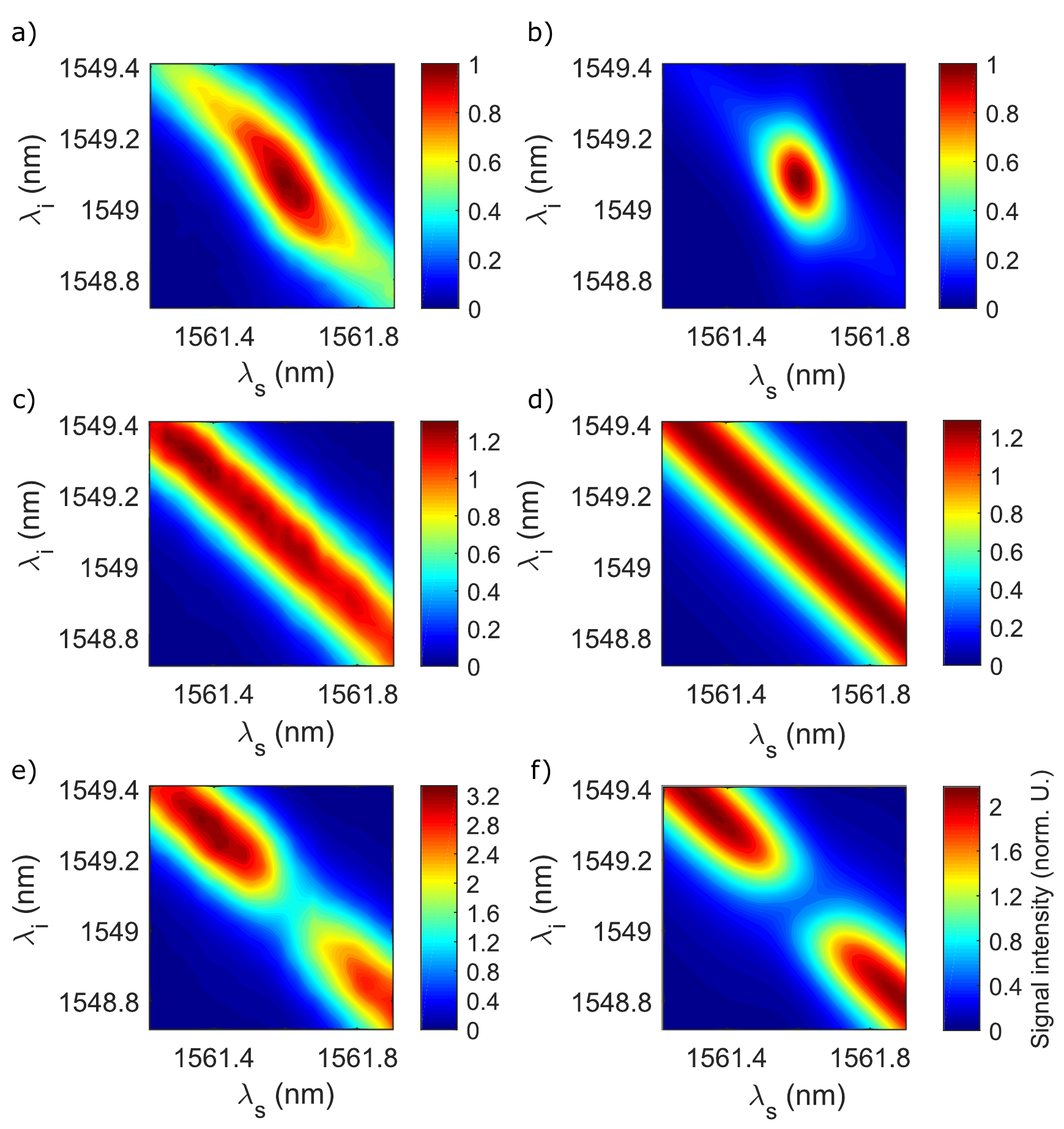}
\caption{(a) Experimental JSI of the resonator. (b) Simulation of the JSI of the resonator. (c) Experimental JSI of the spiral. (d) Simulation of the JSI o the spiral. (d) Experimental interference between the stimulated fields generated by the spiral and the resonator. (e) Simulation of the interference between the stimulated fields.}
\label{fig:2}
\end{figure}

\begin{figure*}[t!]
\centering
\includegraphics[scale = 0.82]{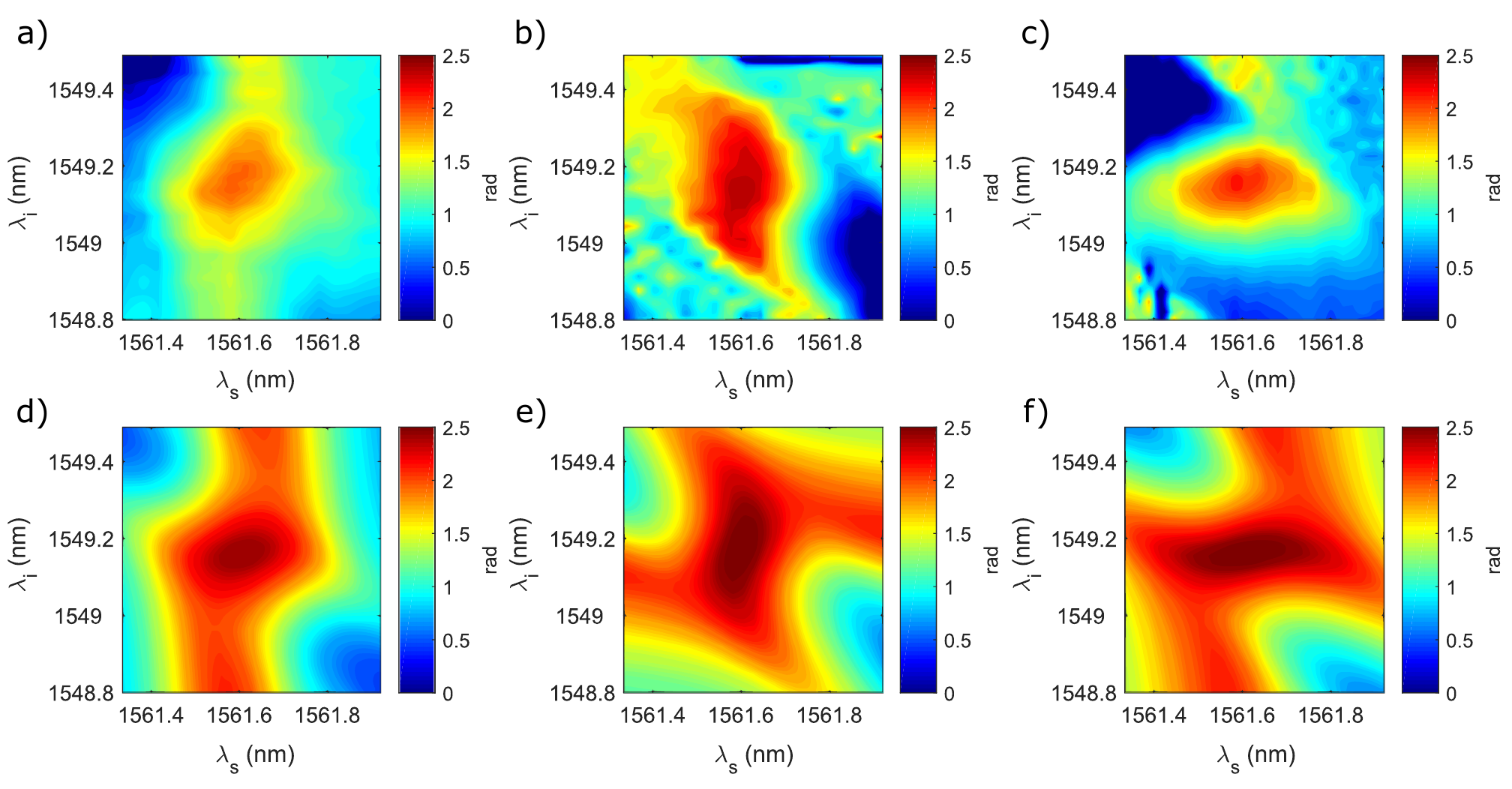}
\caption{(a) Measure of the phase $|\delta|$ obtained using the on-chip filter. (b) Measure of the phase $|\delta|$ obtained using the off-chip filter set to $50\,\textrm{pm}$ of resolution. The Seed laser is scanned across the resonance order $m_s = m_p+1$ (c) Same as in panel (b), but the Seed laser is scanned across the resonance order $m_s = m_p-1$. (d) Simulation of $|\delta|$ taking into account the spectral resolution of the on-chip filter. (e) Simulation of $|\delta|$ taking into account the spectral resolution of the off-chip filter. The Seed laser is scanned across the resonance order $m_s = m_p+1$. (f) Same as in panel (e), but the Seed laser is scanned across the resonance order $m_s = m_p-1$.}
\label{fig:3}
\end{figure*}

The JSI of the resonator and of the spiral are measured by monitoring the output power of the stimulated field when the upper or the lower arm of the device are excited. 
These are shown in Fig.\ref{fig:2}(a,c), while Fig.\ref{fig:2}(b,d) are simulations which uses the same parameters of the experiment. The resolution of the Seed wavelength $\lambda_s$ is $20\,\textrm{pm}$, while the one on the stimulated field is $140\,\textrm{pm}$, and is limited by the linewidth of the on-chip filter. Add-Drop filters with FWHM $<40\,\textrm{pm}$ are routinely available in SOI \cite{borghi2015high}, which can potentially increase the resolution. The calculated fidelities with the simulation are $\textrm{F}=(96.7\pm0.2)$ for the spiral and $\textrm{F}=(91.5\pm0.4)$ for the resonator. Errorbars are computed through Monte Carlo simulations assuming poissonian distribution of the data. The experimental JSI of the resonator reveals to be much broader than expected. This has probably to be attributed to the stimulated radiation generated in the spirals located before (of length $0.65\,\textrm{mm}$) and after (of  length $2.5\,\textrm{mm}$) the resonator. 
\noindent Importantly, in the region where the JSI is more intense, the spurious contribution from the waveguide after the ring is negligible, since both the Pump and the Seed fields are filtered from the resonator. The use of an aMZI filter for pump rejection, after the resonator, could be used to completely suppress this background field. The effect of the relative phase between the resonator and the spiral emerges from Fig.\ref{fig:2}(e,f), which show the interference of the stimulated fields when both sources are excited. The fidelity with the simulation is $\textrm{F}=(95.51\pm0.07)$. In the central region, a gradual suppression of the intensity is observed. The relative phase $\delta$ between the stimulated field of the resonator and the spiral can be obtained from the relation:
\begin{equation}
|\delta| = \arccos \left ( \frac{I_{\textup{int}}-I_{\textup{res}}-I_{\textup{spi}}}{2\sqrt{I_{\textup{res}}I_{\textup{spi}}}} \right )
\label{eq:9}
\end{equation}
where $I_{\textup{int}}$, $I_{\textup{res}}$ and $I_{\textup{spi}}$ refer respectively to the intensity maps in panels (e), (a) and (b) of Fig.\ref{fig:2}. More precisely, $\delta$ is the convolution of the phase of the stimulated field of the resonator with the point spread function of the filter, so that the two coincide only in the limit of an infinitely narrow  bandwidth. 
The phase $|\delta|$ is plotted in Fig.\ref{fig:3}(a), while in Fig.\ref{fig:3}(d) it is compared to simulation.  Figure \ref{fig:3}(a) reveals  that $|\delta|$ is not constant, but has a maximum when $\lambda_s=\lambda_{m_s}$ and $\lambda_i = \lambda_{m_i}$, where $\lambda_{m_{s,i}}$ are the resonance wavelengths of order $m_s$ (Seed) and $m_i$ (Idler). 
According to Eq.\ref{eq:8}, the phase of the stimulated field should be dependent on the seeded resonance order, since $\theta_{\textup{FE}}(\lambda_{s})$ differs from $\theta_{\textup{FE}}(\lambda_{i})$. This is verified by the swapping the Seed laser wavelength from the resonance order $m_s=m_p-1$ to $m_s = m_p+1$, obtaining the phase profiles shown in Fig.\ref{fig:3}(b,c). The external tunable filter is used to increase the resolution. The results are in good agreement with the simulations in Fig.\ref{fig:3}(e,f). The main discrepancies lie outside the main diagonal, and arise from the low counts available in these spectral regions.  A comparison between panels (b) and (c) in Figure \ref{fig:3} clearly shows that $|\delta|$ is not symmetric with respect to the exchange of the seeded resonance, as it would be if $|\delta| \propto \textrm{JSP}$. Even if not shown in Fig.\ref{fig:3}, the same JSI profile is measured in both panels (b) and (c), proving that $I_{\textup{res}}\propto\textrm{JSI}$. This is a remarkable result, since the fact that the system is not seeded in an asymptotic field can be only detected through a phase resolved  measurement. 
\section{Measure of the JSP}
\label{sec:measureJSP}
\begin{figure*}[h!]
\centering
\includegraphics[scale = 0.74]{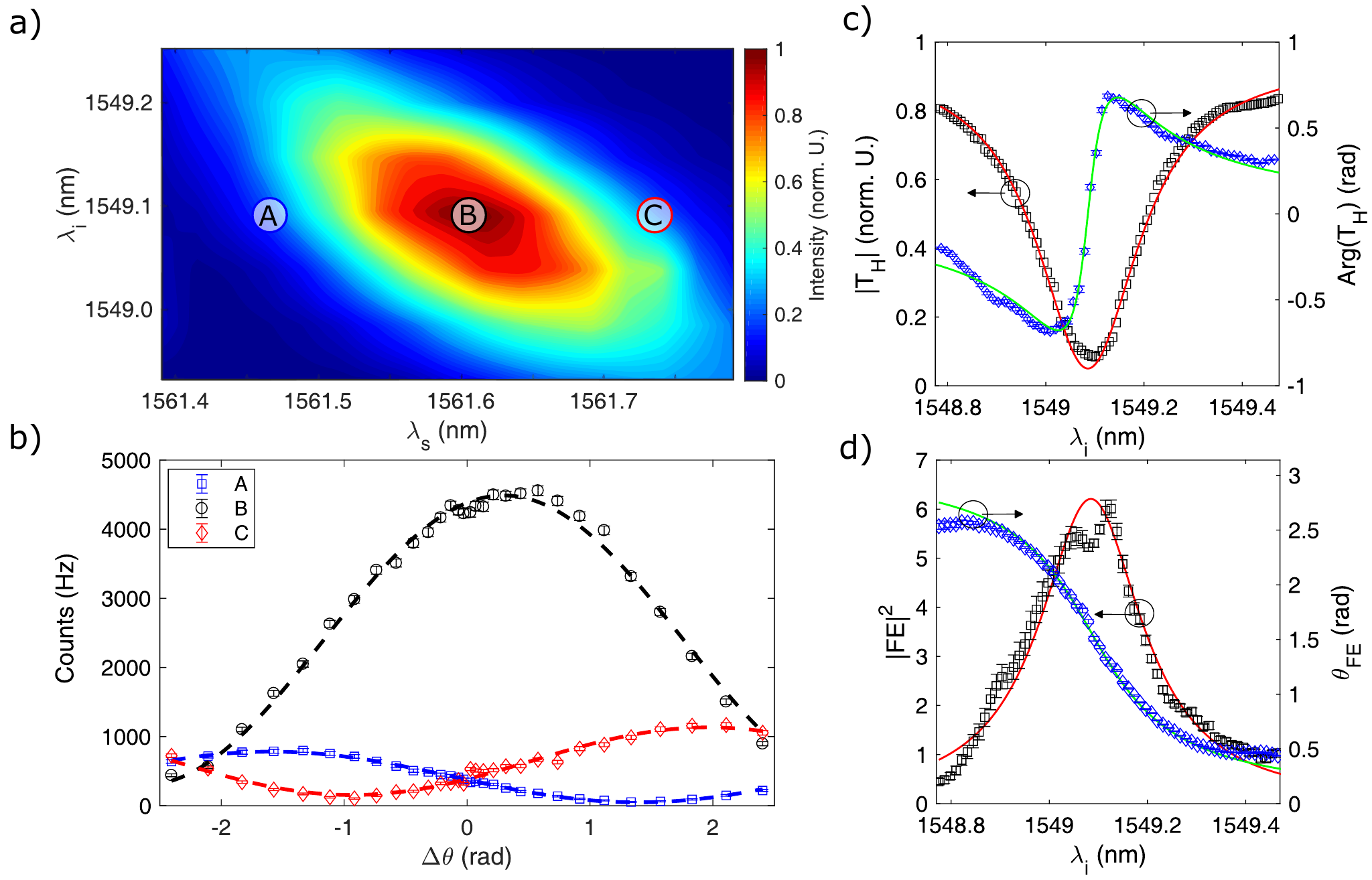}
\caption{(a) Experimental JSI of the resonator in the spectral region where the JSP shown in Fig.\ref{fig:5}(a) has been evaluated. Points labeled by A,B and C refers to the combinations of $(\lambda_s,\lambda_i)$ at which the fringes shown in panel (b) have been recorded. (b) Interference of the stimulated fields of the resonator and the spiral source as a function of the phase $\Delta\theta$. The labels A, B and C refer to the points in the JSI space shown in panel (a). Scatters are from experimental data, solid lines from a fit which uses Eq.\ref{eq:10}. (c) Modulus (black scatters) and phase (blue scatters) of the Through transfer function of the resonator. Solid lines are fit of the experimental data. (d) Modulus square (black scatters) and phase $\theta_{\textup{FE}}$ (blue scatters) of the field enhancement of the resonator. Solid lines are fit of the experimental data.}
\label{fig:4}
\end{figure*}
The phase retrieval method described in Section \ref{sec:measureJSI} does not allow to determine the sign of $\delta$. To this goal, both the resonator and the spiral are coherently excited, and for each combination of $(\lambda_s,\lambda_i)$, the phase $\Delta\theta = \theta_{4}-\theta_{5}$ is scanned. To extract $\delta$, the fringes of their interference pattern $I_{\textup{int}}(\lambda_s,\lambda_i,\Delta\theta)$ are fitted using the relation: 
\begin{equation}
I_{\textup{int}}(\lambda_s,\lambda_i,\Delta\theta) \sim A(\lambda_s,\lambda_i)\cos(\Delta\theta+\delta(\lambda_s,\lambda_i)) +B(\lambda_s,\lambda_i)\label{eq:10}
\end{equation}
\noindent in which $A$ and $B$ are respectively the wavelength dependent amplitude and background of the fringe.
The experimental data is acquired in a spectral grid of $10\times20$ points, and for each, $\Delta\theta$ is varied in $30$ steps. Note that four acquisitions, with $\Delta\theta$ set to $(0,\frac{\pi}{2},\pi,\frac{3\pi}{2})$, would be sufficient to unambiguously determine $\delta$ \cite{jizan2016phase}, but we choose to scan more points to increase the precision.   
Figure \ref{fig:4}(b) reports some of the fringes, which refer to the points labelled as A, B and C in in Fig.\ref{fig:4}(a). In order to improve the fit, the visibility $\textrm{V}$ has been maximized by balancing the intensity of the stimulated fields, yielding an average value of $\textrm{V}=(80.0\pm0.2)\,\%$ without background noise subtraction. The amplitude of the fringes depends on the product of the intensity of the stimulated fields, so it is smaller in the outer points A and C. Due to this fact, in the regions where the JSI of the resonator has negligible intensity, the fringe was too noisy to determine $\delta$. The JSP reconstruction procedure further requires to measure the phase of the field enhancement $\textrm{FE}$ over the Seed resonance. This is done by measuring the complex Through transfer function $T_H$, and by exploiting its relation with the field enhancement,  as detailed in Section \ref{sec:theory}. Modulus and phase of $T_H$ are shown in Fig.\ref{fig:4}(c), while  
the complex field enhancement is shown in Fig.\ref{fig:4}(d).  With both $\delta$ and $\theta_{\textup{FE}}$ in hand, the JSP is reconstructed using Eq.\ref{eq:8}. 
\begin{figure*}[h!]
\centering
\includegraphics[scale = 0.8]{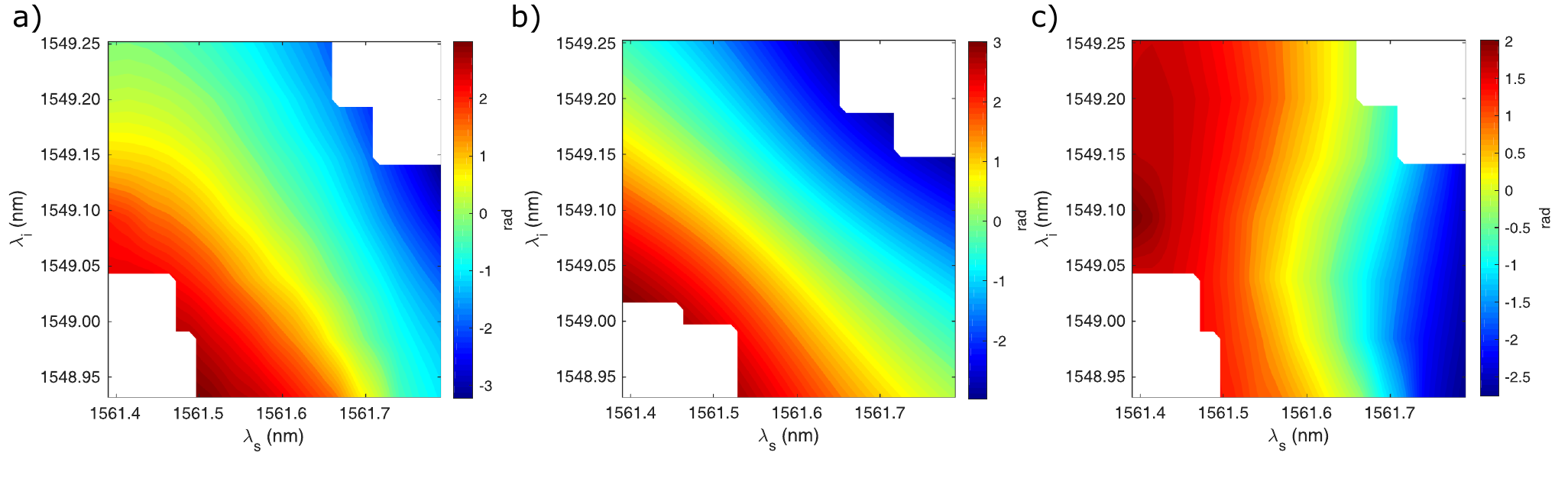}
\caption{(a) Measured JSP of the resonator source. (b) Simulated JSP of the resonator source. (c) Experimental phase $\delta$ of the stimulated field. In the white regions, counts were too low for determining the value of $\delta$.}
\label{fig:5}
\end{figure*}
The measured and the simulated JSPs are compared in Fig.\ref{fig:5}(a,b), and have a fidelity of $\textrm{F}=(83.2\pm0.1)$ with each other. The average error on the phase estimation is $<0.1\,\textrm{rad}$, which mainly arises from the uncertainty on $\delta$. The agreement is better in the central region, where the JSI is more intense and consequently $\delta$ is known with better accuracy. Using only the intensity information of the JSA, shown in Fig.\ref{fig:4}(a), a lower bound for the Schmidt number of $K_{\textup{JSI}}=(1.540\pm0.005)$ is computed.  By including the phase information, this value increases to $K_{\textup{JSP}}=(1.562\pm0.005)$. As a last remark, in Fig.\ref{fig:5}(c) it is reported the measured phase $\delta$ of the stimulated field. Once more, this emphasizes the lack of correspondence with the JSP, which occurs if the corrective phase of the field enhancement is not included in the reconstruction algorithm.  
The total acquisition time for measuring the full complex JSA is about $4$ hours. The speed of the measurement is mainly limited by the use of inefficient grating couplers and by the loss of the external filter, which have a total transmittivity of $\sim 0.02\%$.

\section{Conclusions}
In this paper, it is proposed and experimentally demonstrated a method for reconstructing the complex JSA of an integrated silicon double bus resonator. It is the for the first time, to our knowledge, that the JSA of a resonating source is measured. The approach is based on Stimulated Emission, but removes the need of seeding the system in one of its asymptotic output fields. This is made possible by measuring the complex transfer function of the device, and by exploiting the similarities between the JSAs associated to different output channels. The technique can be extended to single resonators with an arbitrary number of ports, even if these are associated to loss and hence not physically accessible for the reconstruction of the asymptotic output field. 
The JSI and the JSP are entirely measured on a chip, harnessing the compactness, the reconfigurability and the high stability of the optical paths. The number of external resources is minimized, since the scheme eliminates the need of an external reference laser for phase retrieval, being the latter coherently generated on the same chip. The resolution and speed of the measurement could be greatly improved by respectively adopting higher quality factor filters and more efficient grating couplers.
Due to the growing interest in the optimization of integrated resonator sources, either for enhancing their purity or their heralding efficiency, the flexibility allowed by this scheme is expected to become a valuable tool for phase sensitive tomography of a wide class of future devices.

\section*{Acknowledgments}
The author would like to acknowledge prof. M. Liscidini for the support and the useful discussions, I. Farouque, W. McCutcheon, G. Sinclair and S. Paesani for all the fruitful conversations. 

\section*{Funding}
The work was done at the Quantum Engineering and Technology Labs of the University of Bristol, founded by the EPSRC Programme Grant EP/L024020/1. 

\section*{Disclosures}
The author declares no conflicts of interest.

\section*{Appendix A: Hamiltonian tratement of stimulated FWM in double bus resonators}
\label{sec:Hamtheory}
\subsection*{Fields and hamiltonian}
In this appendix, an hamiltonian treatement of stimulated FWM is applied to a double bus (Add-Drop) ring resonator to derive Eq.\ref{eq:8} in the main text. The starting point is the FWM  hamiltonian of Eq.\ref{eq:1}, with the creation and annihilation operators associated to the asymptotic fields sketched in Fig.\ref{fig:1}. 



\noindent Following Ref.\cite{JSPatoms,SET}, the state of the system
$\left|\psi(t)\right\rangle $ at any time $t$ is governed by the
following equation of motion:
\begin{equation}
i\hbar\frac{d\left|\psi(t)\right\rangle }{dt}=U(t)\left|\psi(t)\right\rangle \label{eq:A2}
\end{equation}
\noindent where the evolution operator $U(t)$ is given by:
\begin{equation}
U(t) =  -\sum_{xy}\int S_{xy}(k_{p1},k_{p2},k_{s},k_{i})a_{I,k_{p1}}a_{I,k_{p2}}b_{x,k_{s}}^{\dagger}b_{y,k_{i}}^{\dagger}e^{i\bar{\omega}t}dk_{p1}dk_{p2}dk_{s}dk_{i}+\textrm{h.c.}
\label{eq:A3}
\end{equation}
\noindent where $\bar{\omega} = \omega_{p1}+\omega_{p2}-\omega_{s}-\omega_{i}$. The state $\left|\psi(t)\right\rangle $ is taken of the form:
\begin{equation}
\left|\psi(t)\right\rangle =D_{s}(t)D_{p}(t)D_{i}(t)\left|\textup{mod}(t)\right\rangle \label{eq:A4}
\end{equation}

\noindent where $D_{j}$ is the displacement operator for the $j^{th}$
coherent state. At $t\rightarrow-\infty$, the state is represented
by Pump and Seed pulses which are travelling towards the ring from port I, while at $t\rightarrow\infty$ a coherent
Idler field (as well as the spontaneously generated photon pairs), stimulated from the Pump and the Seed, is outgoing from ports D and T.
The expressions for the displacement operators are given in Eq.\ref{eq:3}.



\noindent At $t\rightarrow-\infty$, the nonlinear interaction has not yet occured,
so it is possible to take as initial conditions $\alpha_{I,k_{p}}(-\infty)=\bar{\alpha}_{I,k_{p}}$,
$\beta_{j,k_{s}}(-\infty)=\bar{\beta}_{j,k_{s}}$, $\gamma_{j,k_{i}}(-\infty)=0$
and $\left|\textup{mod}(-\infty)\right\rangle =\left|0\right\rangle $. 

\noindent The equation of motion for $\left|\textup{mod}(t)\right\rangle $ is obtained 
by differentiating both sides of Eq.\ref{eq:A4}:

\begin{equation}
\frac{d\left|\textup{mod}(t)\right\rangle }{dt}=  \left(O_{s}D_{s}D_{i}D_{p}+D_{s}O_{i}D_{i}D_{p}+D_{s}D_{i}O_{p}D_{p}\right)\left|\textup{mod}(t)\right\rangle+ D_{s}D_{i}D_{p}\frac{d\left|\textup{mod}(t)\right\rangle }{dt}
\label{eq:A8}
\end{equation}

\noindent where $O_{p}=\left(\int\frac{d\alpha_{I,k_{p}}(t)}{dt}a_{I,k_{p}}^{\dagger}dk_{p}-h.c\right)D_{p}$
and $O_{s},O_{i}$ are similarly defined. Using Eq.\ref{eq:A2}, it is possible to rewrite Eq.\ref{eq:A8}
in the form:

\begin{equation}
\frac{d\left|\textup{mod}(t)\right\rangle }{dt}=H_{\textup{eff}}(t)\left|\textup{mod}(t)\right\rangle \label{eq:A9}
\end{equation}

\noindent where the effective hamiltonian $H_{\textup{eff}}$ is given by:
\begin{equation}
H_{\textup{eff}}=  \frac{1}{i\hbar}D_{s}^{\dagger}D_{i}^{\dagger}D_{p}^{\dagger}UD_{s}D_{i}D_{p}-D_{s}^{\dagger}D_{i}^{\dagger}D_{p}^{\dagger} \left(O_{s}D_{s}D_{i}D_{p} \right. \left. D_{s}O_{i}D_{i}D_{p}+D_{s}D_{i}O_{p}D_{p} \right)
\label{eq:A10}
\end{equation}

\noindent If the the Pump, the Signal and the Idler resonances are
confined into three not-overlapping frequency intervals, the different displacement operators commute between each other, i.e., $[D_{j},D_{k}^{\dagger}]_{j\neq k}=0$ and $[D_{j},O_{k}]_{j\neq k}=0$.
\subsection*{Equations of motion }
The effective hamiltonian in Eq.\ref{eq:A10} contains the time derivative of the frequency distributions of the asymptotic input and output fields. In this section their equation of motion are derived. The starting point is the the Heisenberg equation of motion of the Pump operator $a_{I,k}^{\dagger}(t)$, given by \cite{howdoes} :
\begin{equation}
i\hbar\frac{da_{I,k}^{\dagger}(t)}{dt}=[a_{I,k}^{\dagger},V(t)]\label{eq:A11}
\end{equation}
\noindent where the operator $V$ is defined as:
\begin{equation}
V=  -\sum_{xy}\int S_{xy}(k_{p1},k_{p2},k_{s},k_{i})a_{I,k_{p1}}(t)a_{I,k_{p2}}(t)b_{x,k_{s}}^{\dagger}(t)b_{y,k_{i}}^{\dagger}(t)e^{i\bar{\omega}t}dk_{p1}dk_{p2}dk_{s}dk_{i}+\textrm{h.c.}
\label{eq:A11b}
\end{equation}
\noindent Working out the commutator in Eq.\ref{eq:A11} gives:
\begin{equation}
i\hbar\frac{da_{I,k_{p1}}^{\dagger}(t)}{dt}= -2\sum_{xy}\int S_{xy}(k_{p1},k_{p2},k_{s},k_{i})e^{i\bar{\omega}t}a_{I,k_{p2}}(t)b_{x,k_{s}}^{\dagger}(t)b_{y,k_{i}}^{\dagger}dk_{p2}dk_{s}dk_{i}
\label{eq:A12}
\end{equation}

\noindent Similarly, the equation of motion for $b_{j,k_{i}}$
(with $j=\{T,D\}$) can be written as:

\begin{equation}
\begin{aligned}
i\hbar\frac{db_{j,k_{i}}(t)}{dt}= & -\sum_{xy}\int e^{i\bar{\omega}t}a_{I,k_{p1}}(t)a_{I,k_{p2}}(t)\left(S_{yx}(\mathbf{k})b_{y,k_{s}}^{\dagger}(t)\delta_{jx} \right. \\
&\left. +S_{xy}(\mathbf{k})b_{x,k_{s}}^{\dagger}(t)\delta_{jy}\right)dk_{p1}dk_{p2}dk_{s}
\end{aligned}
\label{eq:A13}
\end{equation}

\noindent where the fact that $\mathbf{k} = (k_{p1},k_{p2},k_s,k_i)$ , $[b_{j,k},b_{x,k'}^{\dagger}b_{y,k''}^{\dagger}]=\delta_{jx}\delta(k-k')b_{y,k''}^{\dagger}+\delta_{jy}\delta(k-k'')b_{x,k'}^{\dagger}$ and  $S_{xy}(k_{p1},k_{p2},k_{s},k_{i})=S_{yx}(k_{p1},k_{p2},k_{i},k_{s})$ have been used.
An analogous equation for the operators $b_{j,k_{s}}$ can be obtained by replacing $k_{i}$ with $k_{s}$ in Eq.\ref{eq:A13}. Eq.\ref{eq:A12} and Eq.\ref{eq:A13}, which refer to operators, translates to classical equations of motion for the functions $\alpha_{I,k}(t)$, $\beta_{j,k}(t)$ and $\gamma_{j,k}(t)$ as \cite{howdoes}:
\begin{equation}
i\hbar\frac{d\alpha_{I,k_{p}}}{dt}=2\sum_{xy}\int S_{xy}^{*}(\mathbf{k})e^{-i\bar{\omega}t}\alpha_{I,k_{p2}}^{*}(t)\beta_{x,k_{s}}(t)\beta_{y,k_{i}}dk_{p2}dk_{s}dk_{i}\label{eq:A14}
\end{equation}
\begin{equation}
\begin{aligned}
i\hbar\frac{d\beta_{j,k_{s}}}{dt}= & -\frac{1}{2}\sum_{xy}\int e^{i\bar{\omega}t}\alpha_{I,k_{p1}}(t)\alpha_{I,k_{p2}}(t)\left(S_{xy}(\mathbf{k})\delta_{px}\gamma_{y,k_{i}}^{*}(t)\right. \\
& \left. +S_{yx}(k\mathbf{k})\delta_{jy}\gamma_{x,k_{i}}^{*}(t)\right)dk_{p1}dk_{j2}dk_{i}
\label{eq:A15}
\end{aligned}
\end{equation}
\begin{equation}
\begin{aligned}
i\hbar\frac{d\gamma_{j,k_{i}}}{dt}= & -\frac{1}{2}\sum_{xy}\int e^{i\bar{\omega}t}\alpha_{I,k_{p1}}(t)\alpha_{I,k_{p2}}(t)\left(S_{yx}(\mathbf{k})\delta_{jx}\beta_{y,k_{s}}^{*}(t) \right. \\
& \left. +S_{xy}(\mathbf{k})\delta_{jy}\beta_{x,k_{s}}^{*}(t)\right)dk_{p1}dk_{p2}dk_{s}
\label{eq:A16}
\end{aligned}
\end{equation}

\noindent The factor of $2$ in Eq.\ref{eq:A14} comes from the fact
that the two pumps are degenerate. With Eqs.(\ref{eq:A14}-\ref{eq:A16})
in hand, it is possible to work out the different elements which appear on the
right hand side of Eq.\ref{eq:A10}, obtaining  an explicit expression
for $H_{\textup{eff}}$. After some algebra, the following equalities can be derived:
\begin{equation}
\begin{aligned}
& D_{s}^{\dagger}D_{i}^{\dagger}D_{p}^{\dagger}\left(O_{s}D_{s}D_{i}D_{p}+D_{s}O_{i}D_{i}D_{p}\right)= \\
& -\frac{1}{i\hbar}\sum\int S_{xy}(\mathbf{k})e^{i\bar{\omega}t}\alpha_{I,k_{p1}}\alpha_{k_{p2}}(\beta_{x,k_{s}}^{*}\gamma_{y,k_{i}}^{*}+b_{x,k_{s}}^{\dagger}\gamma_{y,k_{i}}^{*}+b_{x,k_{i}}^{\dagger}\beta_{y,k_{s}}^{*})d\mathbf{k}- \textrm{h.c.}
\label{eq:A17}
\end{aligned}
\end{equation}
\begin{equation}
D_{s}^{\dagger}D_{i}^{\dagger}D_{p}^{\dagger}D_{s}D_{i}O_{p}D_{p}= \frac{2}{i\hbar}\sum\int S_{xy}^{*}(\mathbf{k})e^{-i\bar{\omega}t}\beta_{x,k_{s}}\gamma_{y,k_{i}}(\alpha_{I,k_{p1}}^{*}\alpha_{I,k_{p2}}^{*}+\alpha_{I,k_{p1}}^{*}a_{I,k_{p1}}^{\dagger})d\mathbf{k}-\textrm{h.c.}
\label{eq:A18}
\end{equation}
\begin{equation}
\begin{aligned}
D_{s}^{\dagger}D_{i}^{\dagger}D_{p}^{\dagger}UD_{s}D_{i}D_{p} & = -\sum_{xy}\int S_{xy}(k_{p1},k_{p2},k_{s},k_{i})e^{i\bar{\omega}t}(\alpha_{I,k_{p1}}+a_{I,k_{p1}})
(\alpha_{I,k_{p2}}+a_{I,k_{p2}})\\ &(\beta_{x,k_{s}}^{*}+b_{x,k_{s}}^{\dagger})(\gamma_{y,k_{i}}^{*}+b_{y,k_{i}}^{\dagger})d\mathbf{k}-\textrm{h.c.}
\label{eq:A19}
\end{aligned}
\end{equation}
\noindent where the fact that $D_{j}^{\dagger}a_{k}D_{j}=(\alpha_{k}+a_{k})$ has been used.
\noindent The explicit form of $H_{\textup{eff}}$ is then:
\begin{equation}
\begin{aligned}
H_{\textup{eff}}(t)= & -\frac{1}{i\hbar}\sum_{xy}\int S_{xy}(k_{p1},k_{p2},k_s,k_i)e^{i\bar{\omega}t} \left ( \alpha_{I,k_{p1}}\gamma_{y,k_{i}}^{*}b_{x,k_{s}}^{\dagger}a_{I,k_{p2}} \right. +\\
&  +\alpha_{I,k_{p2}}\gamma_{y,k_{i}}^{*}b_{x,k_{s}}^{\dagger}a_{I,k_{p1}}+a_{I,k_{p1}}a_{I,k_{p2}}\beta_{x,k_{s}}^{*}\gamma_{y,k_{i}}^{*} +\gamma_{y,k_{i}}^{*}b_{x,k_{s}}^{\dagger}a_{I,k_{p1}}a_{I,k_{p2}}+ \\
& +\alpha_{I,k_{p1}}\alpha_{I,k_{p2}}b_{x,k_{s}}^{\dagger}b_{y,k_{i}}^{\dagger} +\alpha_{I,k_{p1}}\beta_{x,k_{s}}^{*}b_{y,k_{i}}^{\dagger}a_{I,k_{p2}}+\gamma_{y,k_{i}}^{*}b_{x,k_{s}}^{\dagger}a_{I,k_{p1}}a_{I,k_{p2}} +\\
& +\alpha_{I,k_{p1}}\alpha_{I,k_{p2}}b_{x,k_{s}}^{\dagger}b_{y,k_{i}}^{\dagger}+\alpha_{I,k_{p1}}\beta_{x,k_{s}}^{*}b_{y,k_{i}}^{\dagger}a_{I,k_{p2}} +\alpha_{I,k_{p1}}b_{x,k_{s}}^{\dagger}b_{y,k_{i}}^{\dagger}a_{I,k_{p2}}+\\ &+\alpha_{I,k_{p2}}\beta_{x,k_{s}}^{*}b_{y,k_{i}}^{\dagger}a_{I,k_{p1}} +\alpha_{I,k_{p2}}b_{x,k_{s}}^{\dagger}b_{y,k_{i}}^{\dagger}a_{I,k_{p1}}+\beta_{x,k_{s}}^{*}b_{y,k_{i}}^{\dagger}a_{I,k_{p1}}a_{I,k_{p2}} +\\
& \left. +b_{x,k_{s}}^{\dagger}b_{y,k_{i}}^{\dagger}a_{I,k_{p1}}a_{I,k_{p2}}+f(t))\right)d\mathbf{k}+\textrm{h.c.}
\end{aligned}
\label{eq:A20}
\end{equation}
\noindent with $f(t)=-2\sum_{xy}\alpha_{I,k_{p1}}(t)\alpha_{I,k_{p2}}(t)\beta_{x,k_{s}}^{*}(t)\beta_{y,k_{i}}^{*}(t)$. Eq.\ref{eq:A20} generalizes the effective hamiltonian in \cite{JSPatoms} to multiple
channels. 
\subsection*{First order solution to the equations of motion}
To first order, the solution of Eq.\ref{eq:A9} is:
\begin{equation}
\left|\textup{mod}(\infty)\right\rangle =\left|0\right\rangle +\intop_{-\infty}^{\infty}H_{\textup{eff}}(t')\left|0\right\rangle dt'\label{eq:A21}
\end{equation}
\noindent Since the only term in $H_{\textup{eff}}$ which does not involve
at least one annihilation operator on the right is $\alpha_{I,k_{p1}}\alpha_{I,k_{p2}}b_{x,k_{s}}^{\dagger}b_{y,k_{i}}^{\dagger}$, it follows that:
\begin{equation}
\left|\textup{mod}(\infty)\right\rangle =\left|0\right\rangle +\sqrt{p_{\textup{tot}}}\sum_{xy}\int\sqrt{\frac{p_{xy}}{p_{\textup{tot}}}}\phi_{xy}(k_{s},k_{i})b_{x,k_{s}}^{\dagger}b_{y,k_{i}}^{\dagger}\left|0\right\rangle dk_{s}dk_{i}\label{eq:A22}
\end{equation}
\noindent where the normalized biphoton wavefunctions
$\phi_{xy}$:
\begin{equation}
\phi_{xy}(k_{s},k_{i})=\frac{2\pi i}{\hbar\sqrt{p_{xy}}}\int S_{xy}(k_{s}+k_{i}-k_{p_{2}},k_{p_{2}},k_{s},k_{i})dk_{p_{2}}\label{eq:A23}
\end{equation}
\noindent have been introduced. The quantity $|\phi_{xy}(k_{s},k_{i})|^{2}dk_{s}dk_{i}$
has then to be interpreted as the probability of finding a photon
with wavevector $k_{s}$ in the asymptotic output state $x$ and a
photon with wavevector $k_{i}$ in the asymptotic output state $y$.
The real numbers $p_{xy}$ and $p_{\textup{tot}}$ are respectively the probability of generating
a photon pair in the channel combination $xy$ and the overall probability
of generating a pair ($p_{\textup{tot}}=\sum_{xy}p_{xy}$). At first order,
the result of Eq.\ref{eq:A21} does not depend on the presence of an
initial Seed beam which stimulates the process. Hence, the final state,
both in the stimulated and in the spontaneous case, is given by:
\begin{equation}
\left|\psi(\infty)\right\rangle =D_{s}(\infty)D_{p}(\infty)D_{i}(\infty)(\left|0\right\rangle +\sqrt{p_{\textup{tot}}}\left|II\right\rangle )\label{eq:24}
\end{equation}
\noindent with $\left|II\right\rangle $ the two-photon state in Eq.\ref{eq:A22}.
\noindent In the absence of a Seed, the solution of Eq.\ref{eq:A16}
is $\gamma_{j,k_{i}}(\infty)=0$, so there is no stimulated
coherent Idler field at the output of the resonator, but only pairs generated by spontaneous FWM. In case that a Seed field is applied, the undepleted Pump approximation gives:
\begin{equation}
\begin{aligned}
\gamma_{j,k_{i}}(\infty)= & \frac{i\pi}{\hbar}\sum_{xy}\int\bar{\alpha}_{k_{s}+k_{i}-k_{p2}}\bar{\alpha}_{k_{p2}}\left(S_{yx}(k_{s}+k_{i}-k_{p2},k_{p2},k_{s},k_{i})\right. \\
&\delta_{jx}\bar{\beta}_{y,k_{s}}^{*} \left. +S_{xy}(k_{s}+k_{i}-k_{p2},k_{p2},k_{s},k_{i})\delta_{jy}\bar{\beta}_{x,k_{s}}^{*}\right)dk_{p2}dk_{s}
\label{eq:A25}
\end{aligned}
\end{equation}
\noindent If the Signal is assumed to be monochromatic at the wavevector
$k_{s0}$, the final expressions for the frequency distributions of
the Idler asymptotic output states are:
\begin{equation}
\gamma_{T,k_{i}}=\sqrt{p_{TT}}\phi_{TT}(k_{s0},k_{i})\beta_{T,k_{s0}}^{*}+\sqrt{p_{TD}}\phi_{TD}(k_{s0},k_{i})\beta_{D,k_{s0}}^{*}\label{eq:A26}
\end{equation}
\begin{equation}
\gamma_{D,k_{i}}=\sqrt{p_{DD}}\phi_{DD}(k_{s0},k_{i})\beta_{D,k_{s0}}^{*}+\sqrt{p_{DT}}\phi_{DT}(k_{s0},k_{i})\beta_{T,k_{s0}}^{*}\label{eq:A27}
\end{equation}
\noindent where the definitions in Eq.\ref{eq:A23} have been used. Equations \ref{eq:A26}-\ref{eq:A27}
are formally equivalent to Eq.\ref{eq:6} in the main text. 
\subsection*{Resonator with an arbitrary number of channels }
The expressions in Eq.\ref{eq:A26}-\ref{eq:A27} are given in terms of the asymptotic output states of the Seed $\beta_{j,k_{s}}$, while
it is much easier to define its initial state in terms of asymptotic
inputs. By using the same formalism of Ref.\cite{asymptotic},
the asymptotic input states $E_{n,k}^{in}$ can be expressed in terms of the asymptotic output ones (and vice-versa) by:
\begin{equation}
E_{n,k}^{in}=\sum_{n'}H_{nn'}^{out}(k)E_{n'k}^{out}\label{eq:A28}
\end{equation}
\begin{equation}
E_{n,k}^{out}=\sum_{n'}H_{nn'}^{in}(k)E_{n'k}^{in}\label{eq:A29}
\end{equation}
\noindent To calculate $H_{nn'}^{in}(k)$, we use the fact that \cite{asymptotic}:
\begin{equation}
E_{n,k}^{in}=E_{n,k}^{iso}+\sum H_{nn'}^{out}(k)E_{n'(-k)}^{iso}\label{eq:A30}
\end{equation}
\noindent where $E_{n'(-k)}^{iso}$ is an outgoing wave from the channel
$n'$ with wavevector $k$. By looking at Fig.\ref{fig:1}, it is evident that:
\begin{equation}
H_{D,A}^{out}=T_{H};\:H_{D,I}^{out}=D_{R};\:H_{T,I}^{out}=T_{H};\:H_{T,A}^{out}=D_{R}\label{eq:A31}
\end{equation}
\noindent where $D_{R}$ and $T_{H}$ are the Drop and Through transfer
functions of the Add-Drop resonator. By using the fact that $H_{nn'}^{in}(k)=\left(H_{nn'}^{out}(k)\right)^{*}$
\cite{asymptotic}:
\begin{equation}
E_{T,k}^{out}=H_{T,I}^{out}(k){}^{*}E_{I,k}^{in}+H_{T,A}^{out}(k){}^{*};\:E_{A,k}^{in}=T_{H}^{*}E_{I,k}^{in}+D_{R}^{*}E_{A,k}^{in}\label{eq:A32}
\end{equation}
\begin{equation}
E_{D,k}^{out}=H_{D,I}^{out}(k){}^{*}E_{I,k}^{in}+H_{D,A}^{out}(k){}^{*};\:E_{A,k}^{in}=D_{R}^{*}E_{I,k}^{in}+T_{H}^{*}E_{A,k}^{in}\label{eq:A33}
\end{equation}
\noindent It is worth to note that the asymptotic output states in Eq.\ref{eq:A32}-\ref{eq:A33} are associated to annihilation operators
\cite{asymptotic}, while the $\beta_{j,k_{s}}$in Eq.\ref{eq:A27}
refer to creation operators (see Eq.\ref{eq:6}). It is possible to shift from
one set to the other by a complex conjugation. By using Eqs.(\ref{eq:A32}-\ref{eq:A33})
into Eqs.(\ref{eq:A26}-\ref{eq:A27}) and considering that in the experiment of this paper $E_{A,k}^{in}=0$:
\begin{equation}
\gamma_{T,k_{i}} =\left(\sqrt{p_{TT}}\phi_{TT}(k_{s0},k_{i})T_{H}^{*}(k_{s0})+\sqrt{p_{TD}}\phi_{TD}(k_{s0},k_{i})D_{R}^{*}(k_{s0})\right)\beta_{I,in}^{*}
\label{eq:A34}
\end{equation}
\begin{equation}
\gamma_{D,k_{i}}=\left(\sqrt{p_{DD}}\phi_{DD}(k_{s0},k_{i})D_{R}^{*}(k_{s0})+\sqrt{p_{DT}}\phi_{DT}(k_{s0},k_{i})T_{H}^{*}(k_{s0})\right)\beta_{I,in}^{*}
\label{eq:A35}
\end{equation}

\noindent The phase matching functions $S_{xy}$ have as explicit
expression \cite{asymptotic}:
\begin{equation}
S_{xy}=\mathbb{N}\int E_{I,k_{p1}}^{in}(\mathbf{r})E_{I,k_{p2}}^{in}(\mathbf{r})\left(E_{x,k_{s}}^{out}(\mathbf{r})\right)^{*}\left(E_{y,k_{i}}^{out}(\mathbf{r})\right)^{*}e^{i\Delta k z}d\mathbf{r_{t}}dz\label{eq:A36}
\end{equation}
\noindent where $\mathbf{r}=(\mathbf{r_{t}},z)$ indicates the spatial coordinates of the integration
($\mathbf{r_{t}}$ is the transverse coordinate along the radial direction
of the ring, while $z$ runs along the ring circumference), $\mathbb{N}$ is a constant which is proportional to the nonlinear susceptibility of the material and $\Delta k = k_{p1}+k_{p2}-k_{s}-k_{i}$. The integration is performed only within the ring due to the higher intensity with respect to the bus waveguides, so
$E_{nk}^{in}\sim \textrm{FE}_{nk}$, where $\textrm{FE}_{nk}$ denotes the internal
field enhancement when the resonator is excited at port $n$.
The relation $E_{nk}^{in}=\left(E_{nk}^{out}\right)^{*}$ 
\cite{asymptotic} can be used to express $S_{xy}$ only in terms of the asymptotic
input fields, so as $S_{xy}\propto E_{I,k_{p1}}^{in}(\mathbf{r})E_{I,k_{p2}}^{in}(\mathbf{r})E_{x,k_{s}}^{in}(\mathbf{r})E_{y,k_{i}}^{in}(\mathbf{r})$.
When the device is symmetric, $\textrm{FE}_{I,k}=\textrm{FE}_{j,k}=\textrm{FE}$
with $j=\{T,D\}$ and consequently $\phi_{xy}=\phi$ where:
\begin{equation}
\phi(k_{s},k_{i})=\frac{2\pi i}{\sqrt{p}\hbar}\mathbb{N}' \textrm{FE}(k_{s})\textrm{FE}(k_{i})\int \textrm{FE}(k_{s}+k_{i}-k_{p_{2}})\textrm{FE}(k_{p_{2}})dk_{p_{2}}\label{eq:A37}
\end{equation}
\noindent in which $\mathbb{N}'$ includes the result of the spatial integration of the asymptotic fields (which gives the inverse of the effective volume) and $e^{i(k_{p1}+k_{p2}-k_{s}-k_{i})z}\sim 1$.
In Eq.\ref{eq:A37}, $p$ represents any of the probabilities $p_{xy}$,
since they all coincide. Equation \ref{eq:A34} reduces to:
\begin{equation}
\gamma_{T,k_{i}}=\sqrt{p}\beta_{I,in}^{*}\phi(k_{s},k_{i})(T_{H}^{*}(k_{s})+D_{R}^{*}(k_{s}))\label{eq:A38}
\end{equation}
\noindent From Temporal Coupled Mode Theory (TCMT) in weak coupling regime \cite{li2016backscattering} it is possible to show that  $D_{R}=i\sqrt{\frac{2}{\tau_{e}}}\textrm{FE}$ and $T_{H}=1+i\sqrt{\frac{1}{\tau_{e}}}\textrm{FE}$,
with:
\begin{equation}
\textrm{FE}=\frac{i\sqrt{\frac{2}{\tau_{e}}}}{\sqrt{\tau_{\textup{rt}}}\left ( \frac{1}{\tau_{\textup{tot}}}-i(\omega-\omega_{j}) \right )}\label{eq:A39}
\end{equation}
\noindent where $\tau_{e}$ is the extrinsic photon lifetime due to the coupling with the bus waveguide, $\tau_{\textup{tot}}=2/\tau_e$ is the total photon lifetime, $\tau_{\textup{rt}}$ is the cavity round-trip time and $\omega_{j}$ is one of the
eigenfrequencies of the resonator (in our case, $j=p,s,i$). The sum
$(T_{H}+D_{R})^{*}$ gives:
\begin{equation}
(T_{H}+D_{R})^{*}=-\frac{\frac{2}{\tau_{e}}-i(\omega-\omega_{j})}{\frac{2}{\tau_{e}}+i(\omega-\omega_{j})}=\frac{\textrm{FE}^{*}}{\textrm{FE}}\label{eq:A40}
\end{equation}
\noindent which inserted into Eq.\ref{eq:A38} gives:
\begin{equation}
\gamma_{T,k_{i}}=\sqrt{p}\beta_{I,in}^{*}\phi(k_{s},k_{i})\left(\frac{\textrm{FE}^{*}}{\textrm{FE}}\right)=\sqrt{p}\beta_{I,in}^{*}|\phi|\exp\left(i(\theta_{\phi}-2\theta_{\textup{FE}})\right)\label{eq:A41}
\end{equation}
\noindent that is the same expression in Eq.\ref{eq:8} of the main manuscript.
In case of a resonator with $M$ channels, in which only one is seeded (arbitrarily this channel can be called the Input), Eq.\ref{eq:A26} generalizes to:
\begin{equation}
\gamma_{T,k_{i}}=\sum_{m=1}^{M}\sqrt{p_{Tm}}\phi_{Tm}(k_{s0},k_{i})H_{m,I}^{out*}(k_{s0})\beta_{I,in}^{*}\label{eq:A42a}
\end{equation}
\noindent From Fig.\ref{fig:1}, it follows  that $H_{m,I}^{out}=T_{H}$ if $m=T$, and $H_{m,I}^{out}=D_{R}^{(m)}$ if $m\neq T$. The function
$D_{R}^{(m)}$ is the Drop transfer function seen from the Input port
of the resonator when the device is excited from the $m^{th}$ port,
and is given by $D_{R}^{(m)}=i\sqrt{\frac{2}{\tau_{e,I}}}\textrm{FE}_{m}$,
where $\tau_{e,I}$ is the extrinsic photon lifetime associated to
the coupling with the Input port. The different field enhancements
$\textrm{FE}_{m}$ can all be expressed relative to the one of the Input port
$\textrm{FE}_{I}$ by $\textrm{FE}_{m}=\sqrt{\frac{\tau_{e,I}}{\tau_{e,m}}}\textrm{FE}_{I}=\sqrt{\frac{\tau_{e,I}}{\tau_{e,m}}}\textrm{FE}_{T}$.
Equations \ref{eq:A36}-\ref{eq:A37} implies that $\phi_{Tm}=\sqrt{\frac{p_{TT}}{p_{Tm}}}\sqrt{\frac{\tau_{e,i}}{\tau_{e,m}}}\phi_{TT}$
so that Eq.\ref{eq:A42a} becomes:
\begin{equation}
\gamma_{T,k_{i}} =  \sqrt{p_{TT}}\phi_{TT}(k_{s0},k_{i})\beta_{I,in}^{*}\left(1+i\textrm{FE}_{T}\left(\sqrt{\frac{2}{\tau_{e,I}}}+\sum_{m\neq T}i\frac{\sqrt{2\tau_{e,I}}}{\tau_{e,m}}\right)\right)^{*}
\label{eq:A43a}
\end{equation}
\noindent Replacing $\textrm{FE}_{T}$ with the expression in Eq.\ref{eq:A39}, yields:
\begin{equation}
\gamma_{T,k_{i}} =  \sqrt{p_{TT}}\phi_{TT}(k_{s0},k_{i})\beta_{I,in}^{*} \left(\frac{-i(\omega-\omega_{s0})+\frac{1}{\tau_{\textup{tot}}}-\left(\frac{2}{\tau_{e,I}}+\sum_{m\neq T}\frac{2}{\tau_{e,m}}\right)}{-i(\omega-\omega_{s0})+\frac{1}{\tau_{\textup{tot}}}}\right)^{*}
\label{eq:A44a}
\end{equation}
\noindent In the numerator of Eq.\ref{eq:A44a}, the term
$\frac{2}{\tau_{e,I}}+\sum_{m\neq T}\frac{2}{\tau_{e,m}}$ equals $\frac{2}{\tau_{\textup{tot}}}$, so that the expression in the parenthesis reduces to the right hand
side of Eq.\ref{eq:A40}, and Eq.\ref{eq:A44a} gets the same formal expression of the two-channel
case in Eq.\ref{eq:A41}. Naturally, one could recover a similar result in a general
channel $m$ by replacing $\phi_{TT}$ in Eq.\ref{eq:A44a}
with $\phi_{mm}$. The information on the number of channels is stored
inside the field enhancement factors.
In all this derivation, no assumptions are made on the nature
of the channels, so Eq.\ref{eq:A44a} holds even in presence
of linear loss. Indeed, as shown in \cite{lossy}, they can be included
by adding a phantom channel in the hamiltonian, which behaves exactly
as any of the other physical channels. 
\section*{Appendix B: Temporal Coupled Mode Theory}
\label{sec:TCMT}
In this section, the expression for the amplitude of the stimulated field in the Add-Drop resonator is derived using TCMT, showing that it agrees with the result of the hamiltonian treatement. The Pump and the Seed fields are modeled as $A_{p}(t)=a_{p}(t)e^{-i\omega_{p}t}$ and $A_{s}(t)=a_{s}(t)e^{-i\omega_{s}t}$, where $a_{p(s)}$ are slowly varying envelopes compared to the carrier frequencies $\omega_{p(s)}$. The equations which govern the (slowly varying) energy amplitudes $u(t)$ inside the resonator are \cite{borghi2019four}:
\begin{equation}
\begin{aligned}
\frac{du_{p}}{dt} = & \left( i(\omega_{p}-\omega_{p0})-\frac{1}{\tau_{p,\textup{tot}}} \right) u_{p}+i\sqrt{\frac{2}{\tau_{p,e}}}a_{p}(t)  \\
\frac{du_{s}}{dt} = & \left( i(\omega_{s}-\omega_{s0})-\frac{1}{\tau_{s,\textup{tot}}}\right)u_{s}+i\sqrt{\frac{2}{\tau_{s,e}}}a_{s}(t) \\
\frac{du_{i}}{dt}= & \left[i(\omega_{i}-\omega_{i0})-\frac{1}{\tau_{i,\textup{tot}}}\right]u_{i}+\gamma u_{p}(t)^{2}u_{s}(t)^{*}
\end{aligned}
\label{eq:t1}
\end{equation}
\noindent where the subscripts $(p,s,i)$ label the Pump, the Seed and the Idler. The nonlinear coupling parameter $\gamma$ is related to the more familiar nonlinear coefficient $\gamma_{\textup{nl}}=\frac{\omega n_{2}}{cA_{\textup{eff}}}$ ($n_2$ is the nonlinear refractive index while $A_{\textup{eff}}$ the effective area of the waveguide) by $\gamma=\frac{\gamma_{\textup{nl}}L_{\textup{res}}}{\tau_{\textup{rt}}^{2}}$,
in which $L_{\textup{res}}$ the resonator perimeter. Consistently with the hamiltonian treatement,  all the parasitic nonlinearities, except FWM, are neglected. Casting the Pump and Seed equations into the Fourier domain and solving for the energy amplitudes gives:
\begin{equation}
U_{p(s)}(\omega)=  \sqrt{\tau_{rt}}\textrm{FE}_{p(s)}(\omega)A_{p(s)}(\omega)
= \frac{i\sqrt{\frac{2}{\tau_{p(s),e}}}}{\frac{1}{\tau_{p(s),\textup{tot}}}-i(\omega-\omega_{p(s)0})}A_{p(s)}(\omega)
\label{eq:t2}
\end{equation}
\noindent where $U_{p(s)}(\omega)=u_{p(s)}(\omega-\omega_{p(s)})$. Inserting Eq.\ref{eq:t2} into Eq.\ref{eq:t1} yields:
\begin{equation}
\begin{aligned}
& U_{i}(\omega) =  \gamma\tau_{\textup{rt}}^{2}\frac{\textrm{FE}_{i}(\omega)}{i\sqrt{\frac{\tau_{i,e}}{2}}}\int \textrm{FE}_{p}(\omega-\omega'-\omega''-\omega_{i}+\omega_{p})\times \\
 & \times\textrm{FE}_{p}(\omega''+\omega_{p})\textrm{FE}_{s}^{*}(\omega'+\omega_{s}) A_{p}(\omega-\omega'-\omega''-\omega_{i}+\omega_{p})A_{p}(\omega''+\omega_{p})A_{s}^{*}(\omega'+\omega_{s})d\omega'd\omega''
\end{aligned}
\label{eq:t3}
\end{equation}
where the convolution property of the Fourier transform is recursively used. The Seed field is assumed to be monochromatic and at frequency $\bar{\omega_{s}}$. In this case, the power amplitude per unit frequency  $P_{i,\textup{res}}=i\sqrt{\frac{\tau_{i,e}}{2}}U_{i}$ of the Idler into the Through (or Drop) waveguide is: 
\begin{equation}
\begin{aligned}
P_{i,\textup{res}}(\omega) & =\gamma_{\textup{nl}}L_{\textup{res}}^{2}\left(\frac{\textrm{FE}(\bar{\omega_{s}})^{*}}{\textrm{FE}(\bar{\omega_{s}})}\right)A_{s}^{*}\textrm{FE}_{i}(\omega)\textrm{FE}_{s}(\bar{\omega_{s}})\\
 & \int \textrm{FE}_{p}(\omega+\bar{\omega_{s}}-\omega')\textrm{FE}_{p}(\omega')A_{p}(\omega+\bar{\omega_{s}}-\omega')A_{p}(\omega')d\omega'
\end{aligned}
\label{eq:t4}
\end{equation}
in which it is used the fact that, from energy conservation, $\omega_{i}=2\omega_{p}-\omega_{s}$.
The fully quantum mechanical calculation for the resonator JSA, $\phi_{\textup{res}}$, gives \cite{vernon2017truly}:
\begin{equation}
\phi(\omega,\bar{\omega}_{s})_{\textup{res}} =  \mathbb{N}_{\textup{res}} \textrm{FE}_{s}(\bar{\omega_{s}})\textrm{FE}_{i}(\omega) \int \textrm{FE}_{p}(\omega+\bar{\omega_{s}}-\omega')\textrm{FE}_{p}(\omega')A_{p}(\omega+\bar{\omega_{s}}-\omega')A_{p}(\omega')d\omega'
\label{eq:t5}
\end{equation}
where $\mathbb{N}_{\textup{res}}$ is a normalization constant. From a comparison between Eq.\ref{eq:t5} and Eq.\ref{eq:t4}, the following identity holds:
\begin{equation}
P_{i,\textup{res}}(\omega)=\mathbb{N}_{\textup{res}}'\left(\frac{\textrm{FE}(\bar{\omega_{s}})^{*}}{\textrm{FE}(\bar{\omega_{s}})}\right)\phi_{\textup{res}}(\omega,\bar{\omega_{s}})
\label{eq:t6}
\end{equation}
in which $\mathbb{N}_{\textup{res}}'$ includes all the pre-factors. The result in Eq.\ref{eq:t6} is formally equivalent to Eq.\ref{eq:8} in the main text and to Eq.\ref{eq:A41}, which has been derived by an hamiltonian treatement. 
\begin{table}[t!]
\centering
\caption{\bf List of all the parameters used to simulate the JSA of the resonator and of the spiral. FEM = Finite Element Method}
\begin{tabular}{cccc}
\hline
Parameter & Value & Source \\
\hline
$L_{\textup{res}}$ & $92.12\,\mu\textrm{m}$ & Experiment  \\
$\tau_{e,p}$ & $24.8$ ps & Experiment \\
$\tau_{e,s}$ & $23.7$ ps & Experiment \\
$\tau_{e,i}$ & $25.9$ ps & Experiment \\
$\tau_{\textup{tot},p}$ & $9.6$ ps & Experiment \\
$\tau_{\textup{tot},s}$ & $9.3$ ps & Experiment \\
$\tau_{\textup{tot},i}$ & $10.0$ ps & Experiment \\
$\lambda_{p0}=2\pi c/\omega_{p0}$ & $1555.32$ nm & Experiment \\
$\lambda_{p}=2\pi c/\omega_{p}$ & $1555.32$ nm & Experiment \\
$\lambda_{s0}=2\pi c/\omega_{s0}$ & $1561.60$ nm & Experiment \\
$\lambda_{i0}=2\pi c/\omega_{i0}$ & $1549.08$ nm & Experiment \\
$\Gamma$ (on chip filter) & $110$ GHz ($140$ pm) & Experiment \\
$\Gamma$ (off chip filter) & $40$ GHz ($50$ pm) & Experiment \\
$L_{\textup{spi}}$ & $2.35\,\textrm{mm}$ & Experiment \\
$A_p$ & / & Fit of laser spectra\\
$L_c$ and $\Delta k$ & / & FEM\\
\hline
\end{tabular}
\label{tab:parameter_simulation}
\end{table}
\subsection*{Appendix C: Simulation of the JSA of the spiral and of the resonator}
\label{sec:simulation_JSA}
The resonator JSA is modeled using Eq.\ref{eq:t5}. The JSI shown in Fig.\ref{fig:2}(b) of the main text takes into account the effect of the finite resolution of the filter for SET. The final expression thus reads:
\begin{equation}
|\phi_{\textup{res}}(\omega_s,\omega_i)|^2 = \int |G(\omega,\omega_i)|^2|\phi_{\textup{res}}(\omega_s,\omega)|^2 d\omega
\label{eq:js0}
\end{equation}
where $G(\omega)$ is the spectral response of the filter. This is modeled as:
\begin{equation}
|G(\omega,\omega_i)|^2=\begin{cases}
\frac{\left(\frac{\Gamma}{2}\right)^{2}}{\left(\frac{\Gamma}{2}\right)^{2}+(\omega-\omega_{i})} & \textrm{on-chip filter}\\
\textrm{rect}\left(\omega\right) & \textrm{off-chip filter}
\end{cases}
\label{eq:js1}
\end{equation}
where $\Gamma$ is the FWHM of the filter, $\omega_i$ is its center frequency and $\textrm{rect}$ is a box-like function. The choice of a Lorentzian lineshape for the on-chip filter comes from the fact that this is implemented using an Add-Drop resonator. The off-chip filter is a tunable fiber Bragg grating, whose measured spectral response well approximates a box-like function. 
The JSA of the spiral is calculated using the following expression \cite{howdoes}:
\begin{equation}
\phi_{\textup{spi}}(\omega_s,\omega_i) =  \mathbb{N}_{\textup{spi}}\int e^{i\Delta kL/2} \textrm{sinc}(L_{\textup{spi}}/L_c) A_p(\omega_s+\omega_i-\omega')A_p(\omega')d\omega'
\label{eq:js3}
\end{equation} 
where $\mathbb{N}_{\textup{spi}}$ is a normalization constant, $L_{\textup{spi}}$ is the spiral length, $\Delta k = k(\omega_s+\omega_i-\omega')+k(\omega')-k(\omega_s)-k(\omega_i)$ is the wavevector mismatch and $L_c=2\pi/\Delta k$ is the coherence length. In the same way of the resonator, the JSI of the spiral plotted in Fig.2(d) in the main text has been convoluted by the spectral response of the filter. 
Table \ref{tab:parameter_simulation} lists all the parameters used in the simulation.

%



\begin{thebibliography}{10}
\newcommand{\enquote}[1]{``#1''}

\bibitem{wang2019integrated}
J.~Wang, F.~Sciarrino, A.~Laing, and M.~G. Thompson, \enquote{Integrated
  photonic quantum technologies,} {\protect\JournalTitle{Nature Photonics}} pp.
  1--12 (2019).

\bibitem{rudolph2017optimistic}
T.~Rudolph, \enquote{Why i am optimistic about the silicon-photonic route to
  quantum computing,} {\protect\JournalTitle{APL Photonics}} \textbf{2}, 030901
  (2017).

\bibitem{politi2008silica}
A.~Politi, M.~J. Cryan, J.~G. Rarity, S.~Yu, and J.~L. OBrien,
  \enquote{Silica-on-silicon waveguide quantum circuits,}
  {\protect\JournalTitle{Science}} \textbf{320}, 646--649 (2008).

\bibitem{harris2016large}
N.~C. Harris, D.~Bunandar, M.~Pant, G.~R. Steinbrecher, J.~Mower, M.~Prabhu,
  T.~Baehr-Jones, M.~Hochberg, and D.~Englund, \enquote{Large-scale quantum
  photonic circuits in silicon,} {\protect\JournalTitle{Nanophotonics}}
  \textbf{5}, 456--468 (2016).

\bibitem{qiang2018large}
X.~Qiang, X.~Zhou, J.~Wang, C.~M. Wilkes, T.~Loke, S.~O\'Gara, L.~Kling, G.~D.
  Marshall, R.~Santagati, T.~C. Ralph \emph{et~al.}, \enquote{Large-scale
  silicon quantum photonics implementing arbitrary two-qubit processing,}
  {\protect\JournalTitle{Nature Photonics}} \textbf{12}, 534 (2018).

\bibitem{wang2018multidimensional}
J.~Wang, S.~Paesani, Y.~Ding, R.~Santagati, P.~Skrzypczyk, A.~Salavrakos,
  J.~Tura, R.~Augusiak, L.~Man{\v{c}}inska, D.~Bacco \emph{et~al.},
  \enquote{Multidimensional quantum entanglement with large-scale integrated
  optics,} {\protect\JournalTitle{Science}} p. eaar7053 (2018).

\bibitem{adcock2019programmable}
J.~C. Adcock, C.~Vigliar, R.~Santagati, J.~W. Silverstone, and M.~G. Thompson,
  \enquote{Programmable four-photon graph states on a silicon chip,}
  {\protect\JournalTitle{Nature communications}} \textbf{10}, 1--6 (2019).

\bibitem{caspani2017integrated}
L.~Caspani, C.~Xiong, B.~J. Eggleton, D.~Bajoni, M.~Liscidini, M.~Galli,
  R.~Morandotti, and D.~J. Moss, \enquote{Integrated sources of photon quantum
  states based on nonlinear optics,} {\protect\JournalTitle{Light: Science \&
  Applications}} \textbf{6}, e17100 (2017).

\bibitem{silverstone2014chip}
J.~W. Silverstone, D.~Bonneau, K.~Ohira, N.~Suzuki, H.~Yoshida, N.~Iizuka,
  M.~Ezaki, C.~M. Natarajan, M.~G. Tanner, R.~H. Hadfield \emph{et~al.},
  \enquote{On-chip quantum interference between silicon photon-pair sources,}
  {\protect\JournalTitle{Nature Photonics}} \textbf{8}, 104 (2014).

\bibitem{vernon2017truly}
Z.~Vernon, M.~Menotti, C.~Tison, J.~Steidle, M.~Fanto, P.~Thomas, S.~Preble,
  A.~Smith, P.~Alsing, M.~Liscidini \emph{et~al.}, \enquote{Truly unentangled
  photon pairs without spectral filtering,} {\protect\JournalTitle{Optics
  letters}} \textbf{42}, 3638--3641 (2017).

\bibitem{grassani2015micrometer}
D.~Grassani, S.~Azzini, M.~Liscidini, M.~Galli, M.~J. Strain, M.~Sorel,
  J.~Sipe, and D.~Bajoni, \enquote{Micrometer-scale integrated silicon source
  of time-energy entangled photons,} {\protect\JournalTitle{Optica}}
  \textbf{2}, 88--94 (2015).

\bibitem{reimer2016generation}
C.~Reimer, M.~Kues, P.~Roztocki, B.~Wetzel, F.~Grazioso, B.~E. Little, S.~T.
  Chu, T.~Johnston, Y.~Bromberg, L.~Caspani \emph{et~al.}, \enquote{Generation
  of multiphoton entangled quantum states by means of integrated frequency
  combs,} {\protect\JournalTitle{Science}} \textbf{351}, 1176--1180 (2016).

\bibitem{faruque2018chip}
I.~I. Faruque, G.~F. Sinclair, D.~Bonneau, J.~G. Rarity, and M.~G. Thompson,
  \enquote{On-chip quantum interference with heralded photons from two
  independent micro-ring resonator sources in silicon photonics,}
  {\protect\JournalTitle{Optics express}} \textbf{26}, 20379--20395 (2018).

\bibitem{silverstone2015qubit}
J.~W. Silverstone, R.~Santagati, D.~Bonneau, M.~J. Strain, M.~Sorel, J.~L.
  O’Brien, and M.~G. Thompson, \enquote{Qubit entanglement between
  ring-resonator photon-pair sources on a silicon chip,}
  {\protect\JournalTitle{Nature communications}} \textbf{6}, 7948 (2015).

\bibitem{liscidini2013stimulated}
M.~Liscidini and J.~Sipe, \enquote{Stimulated emission tomography,}
  {\protect\JournalTitle{Physical review letters}} \textbf{111}, 193602 (2013).

\bibitem{helt2012does}
L.~G. Helt, M.~Liscidini, and J.~E. Sipe, \enquote{How does it scale? comparing
  quantum and classical nonlinear optical processes in integrated devices,}
  {\protect\JournalTitle{JOSA B}} \textbf{29}, 2199--2212 (2012).

\bibitem{zielnicki2018joint}
K.~Zielnicki, K.~Garay-Palmett, D.~Cruz-Delgado, H.~Cruz-Ramirez, M.~F.
  O\'Boyle, B.~Fang, V.~O. Lorenz, A.~B. U\'Ren, and P.~G. Kwiat,
  \enquote{Joint spectral characterization of photon-pair sources,}
  {\protect\JournalTitle{Journal of Modern Optics}} \textbf{65}, 1141--1160
  (2018).

\bibitem{jizan2015bi}
I.~Jizan, L.~Helt, C.~Xiong, M.~J. Collins, D.-Y. Choi, C.~J. Chae,
  M.~Liscidini, M.~Steel, B.~J. Eggleton, and A.~S. Clark, \enquote{Bi-photon
  spectral correlation measurements from a silicon nanowire in the quantum and
  classical regimes,} {\protect\JournalTitle{Scientific reports}} \textbf{5},
  12557 (2015).

\bibitem{eckstein2014high}
A.~Eckstein, G.~Boucher, A.~Lema{\^\i}tre, P.~Filloux, I.~Favero, G.~Leo, J.~E.
  Sipe, M.~Liscidini, and S.~Ducci, \enquote{High-resolution spectral
  characterization of two photon states via classical measurements,}
  {\protect\JournalTitle{Laser \& Photonics Reviews}} \textbf{8}, L76--L80
  (2014).

\bibitem{fang2014fast}
B.~Fang, O.~Cohen, M.~Liscidini, J.~E. Sipe, and V.~O. Lorenz, \enquote{Fast
  and highly resolved capture of the joint spectral density of photon pairs,}
  {\protect\JournalTitle{Optica}} \textbf{1}, 281--284 (2014).

\bibitem{erskine2018real}
J.~Erskine, D.~England, C.~Kupchak, and B.~Sussman, \enquote{Real-time spectral
  characterization of a photon pair source using a chirped supercontinuum
  seed,} {\protect\JournalTitle{Optics letters}} \textbf{43}, 907--910 (2018).

\bibitem{grassani2016energy}
D.~Grassani, A.~Simbula, S.~Pirotta, M.~Galli, M.~Menotti, N.~C. Harris,
  T.~Baehr-Jones, M.~Hochberg, C.~Galland, M.~Liscidini \emph{et~al.},
  \enquote{Energy correlations of photon pairs generated by a silicon microring
  resonator probed by stimulated four wave mixing,}
  {\protect\JournalTitle{Scientific reports}} \textbf{6}, 23564 (2016).

\bibitem{kumar2014controlling}
R.~Kumar, J.~R. Ong, M.~Savanier, and S.~Mookherjea, \enquote{Controlling the
  spectrum of photons generated on a silicon nanophotonic chip,}
  {\protect\JournalTitle{Nature communications}} \textbf{5}, 5489 (2014).

\bibitem{lenzini2018direct}
F.~Lenzini, A.~N. Poddubny, J.~Titchener, P.~Fisher, A.~Boes, S.~Kasture,
  B.~Haylock, M.~Villa, A.~Mitchell, A.~S. Solntsev \emph{et~al.},
  \enquote{Direct characterization of a nonlinear photonic circuits wave
  function with laser light,} {\protect\JournalTitle{Light: Science \&
  Applications}} \textbf{7}, 17143 (2018).

\bibitem{titchener2015generation}
J.~G. Titchener, A.~S. Solntsev, and A.~A. Sukhorukov, \enquote{Generation of
  photons with all-optically-reconfigurable entanglement in integrated
  nonlinear waveguides,} {\protect\JournalTitle{Physical Review A}}
  \textbf{92}, 033819 (2015).

\bibitem{fang2016multidimensional}
B.~Fang, M.~Liscidini, J.~Sipe, and V.~Lorenz, \enquote{Multidimensional
  characterization of an entangled photon-pair source via stimulated emission
  tomography,} {\protect\JournalTitle{Optics express}} \textbf{24},
  10013--10019 (2016).

\bibitem{rozema2015characterizing}
L.~A. Rozema, C.~Wang, D.~H. Mahler, A.~Hayat, A.~M. Steinberg, J.~E. Sipe, and
  M.~Liscidini, \enquote{Characterizing an entangled-photon source with
  classical detectors and measurements,} {\protect\JournalTitle{Optica}}
  \textbf{2}, 430--433 (2015).

\bibitem{jizan2016phase}
I.~Jizan, B.~Bell, L.~Helt, A.~C. Bedoya, C.~Xiong, and B.~J. Eggleton,
  \enquote{Phase-sensitive tomography of the joint spectral amplitude of photon
  pair sources,} {\protect\JournalTitle{Optics letters}} \textbf{41},
  4803--4806 (2016).

\bibitem{ren2012analysis}
C.~Ren and H.~F. Hofmann, \enquote{Analysis of the time-energy entanglement of
  down-converted photon pairs by correlated single-photon interference,}
  {\protect\JournalTitle{Physical Review A}} \textbf{86}, 043823 (2012).

\bibitem{beduini2014interferometric}
F.~A. Beduini, J.~A. Zieli{\'n}ska, V.~G. Lucivero, Y.~A. de~Icaza~Astiz, and
  M.~W. Mitchell, \enquote{Interferometric measurement of the biphoton wave
  function,} {\protect\JournalTitle{Physical review letters}} \textbf{113},
  183602 (2014).

\bibitem{avenhaus2014time}
M.~Avenhaus, B.~Brecht, K.~Laiho, and C.~Silberhorn, \enquote{Time-frequency
  quantum process tomography of parametric down-conversion,}
  {\protect\JournalTitle{arXiv preprint arXiv:1406.4252}}  (2014).

\bibitem{park2017measuring}
K.-K. Park, J.-H. Kim, T.-M. Zhao, Y.-W. Cho, and Y.-H. Kim, \enquote{Measuring
  the frequency-time two-photon wavefunction of narrowband entangled photons
  from cold atoms via stimulated emission,} {\protect\JournalTitle{Optica}}
  \textbf{4}, 1293--1297 (2017).

\bibitem{liscidini2012asymptotic}
M.~Liscidini, L.~Helt, and J.~Sipe, \enquote{Asymptotic fields for a
  hamiltonian treatment of nonlinear electromagnetic phenomena,}
  {\protect\JournalTitle{Physical Review A}} \textbf{85}, 013833 (2012).

\bibitem{vernon2015spontaneous}
Z.~Vernon and J.~Sipe, \enquote{Spontaneous four-wave mixing in lossy microring
  resonators,} {\protect\JournalTitle{Physical Review A}} \textbf{91}, 053802
  (2015).

\bibitem{yang2008spontaneous}
Z.~Yang, M.~Liscidini, and J.~Sipe, \enquote{Spontaneous parametric
  down-conversion in waveguides: a backward heisenberg picture approach,}
  {\protect\JournalTitle{Physical Review A}} \textbf{77}, 033808 (2008).

\bibitem{li2016backscattering}
A.~Li, T.~Van~Vaerenbergh, P.~De~Heyn, P.~Bienstman, and W.~Bogaerts,
  \enquote{Backscattering in silicon microring resonators: a quantitative
  analysis,} {\protect\JournalTitle{Laser \& Photonics Reviews}} \textbf{10},
  420--431 (2016).

\bibitem{borghi2017nonlinear}
M.~Borghi, C.~Castellan, S.~Signorini, A.~Trenti, and L.~Pavesi,
  \enquote{Nonlinear silicon photonics,} {\protect\JournalTitle{Journal of
  Optics}} \textbf{19}, 093002 (2017).

\bibitem{ant}
\url{https://www.appliednt.com}.

\bibitem{borghi2015high}
M.~Borghi, M.~Mancinelli, F.~Merget, J.~Witzens, M.~Bernard, M.~Ghulinyan,
  G.~Pucker, and L.~Pavesi, \enquote{High-frequency electro-optic measurement
  of strained silicon racetrack resonators,} {\protect\JournalTitle{Optics
  letters}} \textbf{40}, 5287--5290 (2015).

\bibitem{JSPatoms}
K.-K. Park, J.-H. Kim, T.-M. Zhao, Y.-W. Cho, and Y.-H. Kim, \enquote{Measuring
  the frequency-time two-photon wavefunction of narrowband entangled photons
  from cold atoms via stimulated emission,} {\protect\JournalTitle{Optica}}
  \textbf{4}, 1293--1297 (2017).

\bibitem{SET}
M.~Liscidini and J.~Sipe, \enquote{Stimulated emission tomography,}
  {\protect\JournalTitle{Physical review letters}} \textbf{111}, 193602 (2013).

\bibitem{howdoes}
L.~G. Helt, M.~Liscidini, and J.~E. Sipe, \enquote{How does it scale? comparing
  quantum and classical nonlinear optical processes in integrated devices,}
  {\protect\JournalTitle{JOSA B}} \textbf{29}, 2199--2212 (2012).

\bibitem{asymptotic}
M.~Liscidini, L.~Helt, and J.~Sipe, \enquote{Asymptotic fields for a
  hamiltonian treatment of nonlinear electromagnetic phenomena,}
  {\protect\JournalTitle{Physical Review A}} \textbf{85}, 013833 (2012).

\bibitem{lossy}
Z.~Vernon and J.~Sipe, \enquote{Spontaneous four-wave mixing in lossy microring
  resonators,} {\protect\JournalTitle{Physical Review A}} \textbf{91}, 053802
  (2015).

\bibitem{borghi2019four}
M.~Borghi, A.~Trenti, and L.~Pavesi, \enquote{Four wave mixing control in a
  photonic molecule made by silicon microring resonators,}
  {\protect\JournalTitle{Scientific reports}} \textbf{9}, 408 (2019).

\end{thebibliography}
\end{document}